\documentclass{article}

\usepackage{microtype}
\usepackage{graphicx}
\usepackage{booktabs} 
\usepackage{multirow} 
\graphicspath{{./}}
\usepackage{tikz}
\usetikzlibrary{decorations.pathreplacing, arrows.meta}

\usepackage{hyperref}

\ifdefined\anonymous
  \usepackage{icml2026}
  \makeatletter
  \renewcommand{\Notice@String}{Under review. Do not distribute.}
  \makeatother
  \hypersetup{pdfsubject={}}
  \renewcommand{\icmlruler}[1]{}
\else
  \usepackage[preprint]{icml2026}
\fi

\usepackage{amsmath}
\usepackage{amssymb}
\usepackage{mathtools}
\usepackage{amsthm}

\usepackage[capitalize,noabbrev]{cleveref}
\usepackage{aliascnt} 

\theoremstyle{plain}

\theoremstyle{definition}

\newaliascnt{assumption}{theorem}
\newtheorem{assumption}[assumption]{Assumption}
\aliascntresetthe{assumption}
\theoremstyle{remark}

\crefname{assumption}{Assumption}{Assumptions}
\Crefname{assumption}{Assumption}{Assumptions}

\icmltitlerunning{Can Power Draw Constrain Covert Compute?}

\begin{document}

\twocolumn[
  \icmltitle{Can Power Draw Constrain Covert Compute? Limits of Analogue Verification for AI Governance}

  \begin{icmlauthorlist}
    \icmlauthor{Tom Kimpson}{unimelb,mats}
    \icmlauthor{Mauricio Baker}{oxford}
    \icmlauthor{Emlyn Graham}{mats}
  \end{icmlauthorlist}

  \icmlaffiliation{unimelb}{School of Mathematics and Statistics, The University of Melbourne, Parkville, VIC, Australia}
  \icmlaffiliation{oxford}{Department of Engineering Science, University of Oxford, Oxford, UK}
  \icmlaffiliation{mats}{Machine Alignment, Transparency \& Security (MATS)}

  \icmlcorrespondingauthor{Tom Kimpson}{tom.kimpson@unimelb.edu.au}

  \icmlkeywords{AI governance, compute verification, power side channel, identifiability, covert compute}

  \vskip 0.3in
]

\printAffiliationsAndNotice{}

\begin{abstract}
Frontier AI treaties or agreements on limiting computation require external verification; an external auditor must be able to confirm how much computation actually ran and that parties are adhering to the agreement.  Analogue, off-chip measurements such as power draw provide an information channel for verification. It is unknown how well these analogue channels can constrain the amount of computation against an adversary who actively tries to subvert the audit. We derive a closed form for $\beta$, the largest hidden computation a power trace cannot exclude, as a fraction of the declared machine capacity. Measurements on NVIDIA A100 GPUs give $\beta = 1.16$ in the worst case, while an adversarial matched-energy schedule hides at least $0.41$. Analogue power measurements alone therefore constrain compute weakly. Additional restrictions granted by the threat model, such as the ability of the verifier to re-execute the declared work at an observed operating point, let the verifier push $\beta$ down to $0.059$ in the maximally restricted case. This gives a quantitative estimate of what analogue measurements can contribute to compute verification.
\end{abstract}

\section{Introduction}
\label{sec:intro}

As frontier AI systems grow more capable
\citep{sevilla2022compute,bengio2026safetyreport}, governance proposals increasingly
rely on \emph{verification}: the ability of an outside party to check what a
monitored operator is doing \citep{reuel2025openproblems,baker2025verifying}. One proposed mechanism is a threshold on compute
\citep{sastry2024computing}. Above a specified number of floating-point
operations (FLOPs), a training run triggers registration, reporting, or audit
obligations \citep{shavit2023chinchilla,baker2025verifying,heim2024thresholds}. For example, U.S.~Executive Order 14110 temporarily required reports on models
trained using more than $10^{26}$ integer or floating-point operations
\citep{whitehouse2023eo14110,whitehouse2025eo14148}.

Compute verification policies need to be robust to adversarial
operators who do not behave honestly, but instead try to subvert the
threshold by hiding computation. Verification methods that rely on passive telemetry such as utilisation counters, on-chip clocks, or memory traffic read silicon the
operator controls.
Consequently they rely on  tamper-resistant read paths and attested
monitoring firmware that do not yet exist
\citep{ogara2025hardware,ansari2026hardware}; even existing telemetry
classifiers lose accuracy against disguised workloads \citep{rahman2026telemetry}.
Similarly, cryptographic proof-of-training protocols
\citep{jia2021proofoflearning,garg2023zkpot} depend on the operator's
cooperation, while active challenge--response schemes \citep{monfared2026timing}
run their probes on the operator's own untrusted hardware.
Zero-knowledge proofs of inference \citep{sun2024zkllm} can be passed
cheaply by a ``hollow'' model \citep{gong2026hollowllm}, while
proof-of-useful-work schemes \citep{pearl2026pouw} provide lower
bounds on the declared work only.

``Off-chip'' analogue sensors have been suggested as an alternative verification method
\citep{baker2025verifying}. This involves measurement hardware physically separate from
the monitored chips such as a meter on the power feed, or an antenna, a microphone, or a thermal camera beside the
rack. Each
reads a physical side channel such as power, electromagnetic, acoustic, or
thermal emissions and so its measurements reach the verifier without passing
through any silicon or software the operator controls. Such sensors could
be retrofitted to existing data centres and physically secured against
tampering by the adversary.

There are reasons for optimism: in non-adversarial settings, physical
emissions are known to leak workload information. Physical side channels, such as a GPU's
magnetic emanations \citep{maia2022magnetic} and snooped memory-bus traffic
\citep{hu2020deepsniffer}, let an observer recover model architectures and
hyperparameters. Software-visible power and energy counters
\citep{gao2023deeptheft,chaudhuri2025energon}, and other counters and timing
\citep{naghibijouybari2018rendered,wei2020leaky}, let a co-resident process
do the same. On-node telemetry
and on-chip counters classify workloads
\citep{tang2022supercloud,xu2026wave,gangwal2019cryptomining}, and power
draw betrays anomalous computation \citep{clark2013wattsupdoc} and running
GPU activity \citep{arefin2024hdmi}. A related body of work characterises
power draw for GPUs and data centres
\citep{patel2023polca,choukse2025power,latif2025power}. However, these are forward models, fitted largely in cooperative conditions, i.e. they map known
workloads to their emissions. Verification poses the inverse
problem (i.e. emissions to workload) against an operator actively trying to
evade the sensor. It is an open question how well analogue-sensor
measurements can constrain compute in that setting. This paper addresses that
question for power draw measurements, asking \emph{how much} computation ran
rather than what type of workload. The power channel is the channel most directly linked to
computation, since every operation costs energy and total energy therefore
budgets total work. We defer the study of other channels to future work.

This paper is organised as follows. \Cref{sec:setup} sets up the problem mathematically.
\cref{sec:floor} derives a conservative upper bound on the covert compute a
power trace admits, in closed form, from a handful of physical bands.
\Cref{sec:howbig} empirically evaluates that bound by measuring the
bands directly on NVIDIA A100 hardware.
\Cref{sec:decreasing-beta} shows how additional information gained by the
verifier beyond the power trace (e.g.\ a precision declaration) pushes the bound lower. We close with a discussion in
\cref{sec:discussion}.

\section{Setup and threat model}
\label{sec:setup}

\subsection{The forward model}
\label{subsec:forward}

Fix an observation window $t \in [0,T]$ and let $F(t)$ be the instantaneous
rate of arithmetic operations, bounded by the hardware capacity
$F_{\max}$. The forward model of $F(t)$ to power follows from the CMOS power law
\citep{chandrakasan1992lowpower}; chip power splits into a switching term and
a static term,
\begin{equation}
	P \;=\; \underbrace{\alpha\,C\,V^2 f}_{\text{dynamic}} \;+\; P_{\mathrm{stat}},
	\label{eq:cmoslaw}
\end{equation}
with $V$ the supply voltage, $f$ the clock frequency, $C$ the total capacitance of
the logic, and $\alpha$ the activity factor, i.e. the fraction of $C$ that
actually toggles in a given cycle ($0 \le \alpha \le 1$), so that
$\alpha C$ is the capacitance switched per clock tick. A datapath that retires $N_{\mathrm{op}}$
nominal operations per cycle runs at operation rate $F = N_{\mathrm{op}}f$.
Substituting $f = F/N_{\mathrm{op}}$ and collecting constants gives the
forward model
\begin{equation}
	P(t) \;=\; r(t)\,F(t) \;+\; P_0(t), \qquad
	r \;\equiv\; \alpha\,\Bigl(\frac{C}{N_{\mathrm{op}}}\Bigr) V^2 ,
	\label{eq:forward}
\end{equation}
where $P_0$ collects $P_{\mathrm{stat}}$ and the remaining non-compute
overhead (idle draw, cooling, I/O). We call $r$ the \emph{exchange rate}
[J/op]. Note that the clock frequency
has dropped out: what an operation costs does not depend on
how fast the operations run, only on what each switching event costs and how
many operations it delivers. \Cref{app:cmos} gives this derivation in full.

The exchange rate $r(t)$ is not a constant of the device but a property of the
execution. The activity factor $\alpha$ depends on the operand values being
multiplied \citep{chandrakasan1992lowpower}. The capacitance switched per operation,
$C/N_{\mathrm{op}}$, depends on the numeric format (which sets both the
datapath width and the packing $N_{\mathrm{op}}$) and on where the operands
reside \citep{horowitz2014computing}. The supply voltage $V$ tracks the
voltage--frequency operating point \citep{pallipadi2006ondemand}. Every one
of them is under the prover's control, and none is visible to an off-chip
meter. Empirically the spread is wide: component-level analyses of a 45\,nm
process put arithmetic and on-chip memory access at $0.03$--$100$\,pJ per
operation depending on format and operand locality
\citep{horowitz2014computing}. 

In practice, the meter reports only a noisy version of the power,
\begin{equation}
	P_{\mathrm{obs}}(t) = P(t) + \eta(t),
	\label{eq:obs}
\end{equation}
with $\eta(t)$ a zero-mean measurement noise the prover cannot forge. The
noise is a measurement limit; zero-mean noise with a suitable correlation
structure (stationary, finitely correlated) averages away over
the window and shrinks the longer the verifier watches. Throughout this paper
we therefore assume a perfect meter and work with the clean power $P(t)$.

We count
hardware operations (i.e. what the silicon actually executes) rather than
model FLOPs (i.e. the algorithmically required operations).  Every executed operation costs joules, so
the meter sees hardware operations. We count them as
nominal operations of whichever numeric format runs. For example a multiply--add
retires two nominal ops whether it executes at fp32 or int8, and integer
multiply--adds are ops, not FLOPs. 

\subsection{The threat model}
\label{subsec:threat}

There are two parties. The \emph{prover} is the monitored operator. It controls the
workload, the operand values, the numeric formats, the clocks, and the idle
states of the machine. It may wish to run compute it does not declare.
The \emph{verifier} is the inspecting party: it reads $P_{\mathrm{obs}}$ from
an off-chip meter it trusts, and it never sees $r$, $F$, or $P_0$ separately.
The prover declares a workload of $C_{\mathrm{dec}}$ operations. The verifier wants to know if  $C_{\mathrm{dec}}$ is consistent with $P_{\mathrm{obs}}$. The verifier makes two assumptions:
\begin{assumption}[Trusted meter]
\label{ass:meter}
The power trace cannot be forged: $P_{\mathrm{obs}}$ differs from the true
draw $P$ only by zero-mean noise, which we set to zero throughout.
\end{assumption}

\begin{assumption}[Trusted declaration]
\label{ass:declaration}
The declared computation really ran: the smallest total consistent with the
window is $C_{\mathrm{dec}}$ itself. A power meter cannot by itself confirm
this; independently checking that the declared work ran is a separate
verification problem---the domain of re-execution, proof-of-learning,
zero-knowledge proof and proof-of-useful-work protocols
\citep{jia2021proofoflearning,garg2023zkpot,sun2024zkllm,pearl2026pouw}.
\end{assumption}
Everything else is under the prover's control and unobserved: the exchange
rate $r(t)$, the overhead $P_0(t)$, and how the operation rate $F(t)$ splits
between declared and covert work.

\subsection{The inverse model}
\label{subsec:noninjective}

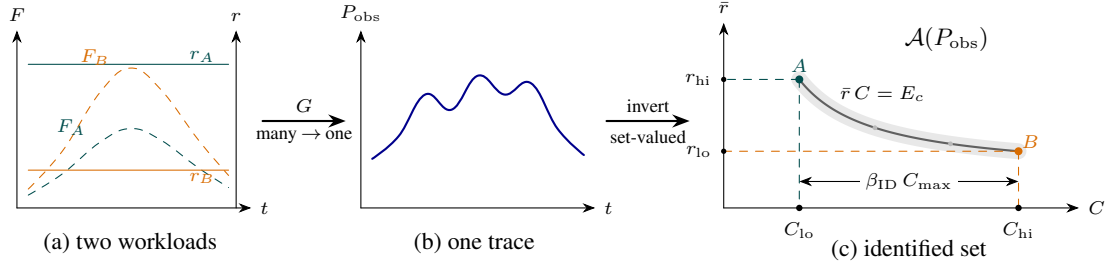
\begin{figure*}[t]
	\centering
	\begin{tikzpicture}[>=Stealth, line join=round, line cap=round]

		\begin{scope}[shift={(0,0)}]
			\draw[->] (0,0.55) -- (3.1,0.55) node[right] {\scriptsize $t$};
			\draw[->] (0,0.55) -- (0,2.9) node[above] {\scriptsize $F$};
			\draw[->] (2.9,0.55) -- (2.9,2.9) node[above] {\scriptsize $r$};
			\draw[teal!60!black] (0.15,2.45) -- (2.8,2.45);
			\draw[teal!60!black, dashed] plot[smooth, tension=0.7] coordinates
			{(0.15,0.72) (0.6,1.0) (1.1,1.42) (1.5,1.6) (1.9,1.45) (2.4,1.08) (2.8,0.82)};
			\draw[orange!85!black] (0.15,1.05) -- (2.8,1.05);
			\draw[orange!85!black, dashed] plot[smooth, tension=0.7] coordinates
			{(0.15,0.8) (0.6,1.3) (1.1,2.08) (1.5,2.4) (1.9,2.12) (2.4,1.45) (2.8,0.98)};
			\node[teal!60!black, anchor=south, inner sep=0.5pt] at (2.45,2.46) {\scriptsize $r_A$};
			\node[orange!85!black, anchor=north, inner sep=0.5pt] at (2.45,1.04) {\scriptsize $r_B$};
			\node[teal!60!black, anchor=east, inner sep=1pt] at (0.95,1.62) {\scriptsize $F_A$};
			\node[orange!85!black, anchor=south east, inner sep=0.5pt] at (1.25,2.42) {\scriptsize $F_B$};
			\node[anchor=north] at (1.5,0.35) {\footnotesize (a) two workloads};
		\end{scope}

		\draw[->, thick] (3.25,1.7) -- (4.35,1.7);
		\node[anchor=south, inner sep=1pt] at (3.8,1.78) {\scriptsize $G$};
		\node[anchor=north, inner sep=1pt] at (3.8,1.62) {\scriptsize many$\,\to\,$one};

		\begin{scope}[shift={(4.55,0)}]
			\draw[->] (0,0.55) -- (3.1,0.55) node[right] {\scriptsize $t$};
			\draw[->] (0,0.55) -- (0,2.9) node[above] {\scriptsize $P_{\mathrm{obs}}$};
			\draw[thick, blue!55!black] plot[smooth, tension=0.7] coordinates
			{(0.15,1.2) (0.5,1.5) (0.85,2.05) (1.2,1.85) (1.55,2.3)
				(1.9,2.05) (2.25,2.2) (2.6,1.6) (2.95,1.25)};
			\node[anchor=north] at (1.5,0.35) {\footnotesize (b) one trace};
		\end{scope}

		\draw[->, thick] (7.8,1.7) -- (8.9,1.7);
		\node[anchor=south, inner sep=1pt] at (8.35,1.78) {\scriptsize invert};
		\node[anchor=north, inner sep=1pt] at (8.35,1.62) {\scriptsize set-valued};

		\begin{scope}[shift={(9.35,0)}]
			\draw[->] (0,0.55) -- (4.7,0.55) node[right] {\scriptsize $C$};
			\draw[->] (0,0.55) -- (0,3.0) node[above] {\scriptsize $\bar r$};
			\draw[line width=6pt, gray!18, line cap=round, smooth, samples=60, domain=1.0:3.9]
			plot (\x,{1.2776/\x+0.9724});
			\draw[gray!75!black, thick, smooth, samples=60, domain=1.0:3.9]
			plot (\x,{1.2776/\x+0.9724});
			\node[anchor=south west, inner sep=1pt] at (1.5,1.92) {\scriptsize $\bar r\,C=E_c$};
			\node[anchor=south] at (2.95,2.5) {\footnotesize $\mathcal{A}(P_{\mathrm{obs}})$};
			\fill[gray!55] (2.0,1.611) circle (0.9pt);
			\fill[gray!55] (3.0,1.398) circle (0.9pt);
			\fill[teal!60!black]   (1.0,2.25) circle (1.5pt) node[above, inner sep=2.5pt] {\scriptsize $A$};
			\fill[orange!85!black] (3.9,1.3)  circle (1.5pt) node[above right, inner sep=1pt] {\scriptsize $B$};
			\draw[dashed, teal!60!black]   (1.0,2.25) -- (0,2.25);
			\draw[dashed, orange!85!black] (3.9,1.3)  -- (0,1.3);
			\fill (0,2.25) circle (1.0pt) node[left=1pt] {\scriptsize $r_{\mathrm{hi}}$};
			\fill (0,1.3)  circle (1.0pt) node[left=1pt] {\scriptsize $r_{\mathrm{lo}}$};
			\draw[dashed, teal!60!black]   (1.0,2.25) -- (1.0,0.55);
			\draw[dashed, orange!85!black] (3.9,1.3)  -- (3.9,0.55);
			\fill (1.0,0.55) circle (1.2pt) node[below=2pt] {\scriptsize $C_{\mathrm{lo}}$};
			\fill (3.9,0.55) circle (1.2pt) node[below=2pt] {\scriptsize $C_{\mathrm{hi}}$};
			\draw[<->] (1.0,0.9) -- (3.9,0.9) node[midway, fill=white, inner sep=2pt] {\scriptsize $\beta_{\mathrm{ID}}\,C_{\max}$};
			\node[anchor=north] at (2.45,0.25) {\footnotesize (c) identified set};
		\end{scope}

	\end{tikzpicture}
	\caption{Non-injectivity of the forward map. (a)~Two workloads ($A$
		little compute at a high rate, $B$ much compute at a low rate) whose
		products $rF$ coincide, so (b)~they drive the identical power trace.
		(c)~Inverting recovers only the set of consistent explanations
		$\mathcal{A}(P_{\mathrm{obs}})$: the rate band
		$[r_{\mathrm{lo}},r_{\mathrm{hi}}]$ cuts the constant-energy arc
		$A$--$B$, whose projection onto the $C$-axis spans
		$[C_{\mathrm{lo}},C_{\mathrm{hi}}]$; relative to capacity that width is
		the physical ambiguity $\beta_{\mathrm{ID}}$. The faint grey
		ribbon is meter noise (ignored in this work). Each rate is drawn
		constant only for clarity; in reality $r(t)$ varies in time. This does
		not matter for a question about total operations, which depends on $r(t)$
		only through its op-weighted mean $\bar r = E_c/C$ (compute energy
		$E_c = \int_0^T rF\,dt$ over total operations $C$). That mean is panel (c)'s
		vertical axis, so letting $r(t)$ vary leaves the picture unchanged.}
	\label{fig:noninjective}
\end{figure*}

Define $G$ as the forward map that carries a workload to its power trace,
\begin{equation}
	G : \bigl(F(\cdot),\, r(\cdot),\, P_0(\cdot)\bigr) \;\longmapsto\; P(\cdot),
	\label{eq:operator}
\end{equation}
defined pointwise by \cref{eq:forward}. The verifier faces an inverse
problem: read the power trace and recover the workload that produced it,
i.e.\ invert $G$. But $G$ is non-injective. Since $r(\cdot)$ is free
to vary, a low rate spent on much computation draws exactly the same power as
a high rate spent on little. Distinct workloads $(F,r,P_0)$ map to identical
traces, and a many-to-one map has no inverse. Inversion therefore returns the identified set $\mathcal{A}(P_{\mathrm{obs}})$ which
collects every physically admissible workload that reproduces
$P_{\mathrm{obs}}$ (\cref{fig:noninjective}).

The question \emph{``how many operations ran over the window?''} corresponds to one
specific projection of $\mathcal{A}(P_{\mathrm{obs}})$. Define the total
compute
\begin{equation}
	C \;=\; \int_0^T F\,dt \quad [\text{op}].
	\label{eq:totalcompute}
\end{equation}
Reading $C$ off each admissible workload projects the identified set onto the
$C$-axis,
\begin{equation}
	\bigl\{\, C \;:\; (F,r,P_0) \in \mathcal{A}(P_{\mathrm{obs}}) \,\bigr\} \;=\;
	[\,C_{\mathrm{lo}},\, C_{\mathrm{hi}}\,].
	\label{eq:projection}
\end{equation}
We define the physical ambiguity $\beta_{\mathrm{ID}}$ as the width of
that interval, measured against the
machine's peak declared-format capacity $C_{\max} = F_{\max}^{\mathrm{dec}}T$,
\begin{equation}
	\beta_{\mathrm{ID}} \;\equiv\; \frac{C_{\mathrm{hi}} - C_{\mathrm{lo}}}{C_{\max}}.
	\label{eq:betadef}
\end{equation}
Under \cref{ass:declaration} the lower edge is the declaration itself,
$C_{\mathrm{lo}} = C_{\mathrm{dec}}$.
The interval runs from that honest baseline up to
the most compute the meter cannot rule out, so the upper edge is the declared
work plus the largest covert load a cheater could slip past the meter,
\begin{equation}
	C_{\mathrm{hi}} \;=\; C_{\mathrm{dec}} \;+\; C_{\mathrm{cov}}^{\max}.
	\label{eq:chi}
\end{equation}
Substituting $C_{\mathrm{lo}} = C_{\mathrm{dec}}$ and \cref{eq:chi} into
\cref{eq:betadef} leaves a single unknown,
\begin{equation}
	\beta_{\mathrm{ID}} \;=\; \frac{C_{\mathrm{hi}} - C_{\mathrm{lo}}}{C_{\max}}
	\;=\; \frac{C_{\mathrm{cov}}^{\max}}{C_{\max}}.
	\label{eq:beta-reduced}
\end{equation}
The width $C_{\mathrm{hi}} - C_{\mathrm{dec}} = C_{\mathrm{cov}}^{\max}$ is
exactly the covert operations a cheater could hide. A wider interval means a larger
$\beta_{\mathrm{ID}}$ and more room for undeclared work.
The unit $C_{\max}$ counts operations in the \emph{declared} format. Covert
work may run in a faster format, for example int8 against declared fp16
(\cref{subsec:capclip}), so $\beta_{\mathrm{ID}}$ can exceed one. A value
above one means the meter cannot exclude more hidden work than the machine
could do in a full window of declared-format operations.

To compute $C_{\mathrm{cov}}^{\max}$ exactly, we would need the set of
schedules that a real device can run. We do not know that set. Instead, we
measure a \emph{band} for each quantity the verifier cannot see: the range,
from lowest to highest value, that the quantity takes in our experiments.
There are three bands. They are the energy cost per operation of declared
work, the energy cost per operation of covert work, and the overhead energy
(\cref{sec:floor}). We then put each quantity at the edge of its band that
helps the cheater most, independently of the other two. The resulting width
is the \emph{relaxed bound} $\beta$. Like $\beta_{\mathrm{ID}}$, it is a
single number in units of $C_{\max}$. It can only be wider:
$\beta_{\mathrm{ID}} \leq \beta$. The two are equal only if one schedule can
reach all three chosen edges at the same time (\cref{app:boundkind}). For a
security claim this is the safe direction: $\beta$ never understates the
hidden compute. When we read $\beta$ as
a statement about what the hardware can actually schedule, the difference
matters.

\section{Derivation of \texorpdfstring{$\beta$}{beta}}
\label{sec:floor}

We now derive a closed form for the relaxed bound $\beta$. Two physical constraints cap the
covert work that can run inside the observation window $[0,T]$. The first is the
\emph{energy budget}: the total the meter reports must stay explainable as
honest declared work, so covert operations can only spend joules that
ambiguity leaves free (\cref{subsec:derivation}). The second is
\emph{capacity}: covert operations must also fit on the machine, in whatever
machine-time the declared work leaves unused (\cref{subsec:capclip}). The
bound $\beta$ is the smaller of the two caps (\cref{subsec:betafloor}).

\subsection{The energy budget}
\label{subsec:derivation}

The derivation of the first constraint is straightforward: integrate the forward
model over the window, split the total energy into declared, covert, and
overhead pieces, and ask how large the covert piece can grow before the total
stops looking honest. We work with the integrated trace (i.e. the total
energy) rather than the resolved time series: treating power as affine in the
operation rate (\cref{eq:forward}) lets its integral depend on the workload
only through time-integrated quantities.

Integrating the forward model \cref{eq:forward} over $[0,T]$, the left-hand
side is the total energy the meter reports and the right-hand side falls
into two pieces,
\begin{equation}
	E \;=\; \underbrace{\int_0^T r(t)\,F(t)\,dt}_{\text{compute energy } E_c}
	\;+\; \underbrace{\int_0^T P_0(t)\,dt}_{\text{overhead } E_0}.
	\label{eq:integrated}
\end{equation}
The overhead $E_0$ is a nuisance the verifier knows only within a band
$E_0 \in [E_{0,\mathrm{lo}}, E_{0,\mathrm{hi}}]$. We do not ``subtract
off'' a known baseline: the prover controls fans, clocks, and idle states,
so a known-and-removed $P_0$ would hand it a hiding channel.

The compute energy $E_c$ splits into a declared block and a covert block,
\begin{multline}
	\int_0^T r(t)\,F(t)\,dt \;=\;\\
	\underbrace{\int_0^T r^{\mathrm{dec}}(t)\,F_{\mathrm{dec}}(t)\,dt}_{\text{declared}} \;+\;
	\underbrace{\int_0^T r^{\mathrm{cov}}(t)\,F_{\mathrm{cov}}(t)\,dt}_{\text{covert}}.
	\label{eq:dec-cov-split}
\end{multline}
The declared block runs $C_{\mathrm{dec}} = \int_0^T F_{\mathrm{dec}}\,dt$
operations, taken known, and the covert block runs the excess
$C_{\mathrm{cov}} = C - C_{\mathrm{dec}}$; this is the compute we are trying
to bound. Write each block's energy as its operation count times a single mean
cost per operation, i.e.\ the op-weighted averages
\begin{equation}
	\bar r^{\,i} \equiv \frac{1}{C_{i}}\int_0^T r^{i}(t)\,F_{i}(t)\,dt
	\;=\; \int_0^T r^{i}(t)\,\frac{F_{i}(t)}{C_{i}}\,dt,
	\label{eq:barr}
\end{equation}
for $i \in \{\mathrm{dec},\,\mathrm{cov}\}$. The second form makes the
averaging explicit: the weight $F_{i}(t)\,dt/C_{i}$ is the fraction of the
block's operations that run in $[t, t+dt]$; it is nonnegative and integrates to
one. The machine therefore draws
\begin{equation}
	E \;=\; \underbrace{\bar r^{\mathrm{dec}}\, C_{\mathrm{dec}}}_{\text{declared work}}
	\;+\; \underbrace{\bar r^{\mathrm{cov}}\, C_{\mathrm{cov}}}_{\text{covert work}}
	\;+\; \underbrace{E_0}_{\text{overhead}},
	\label{eq:cheaterbudget}
\end{equation}
and the verifier sees only $E$.

The two mean costs are bounded,
$\bar r^{\mathrm{dec}} \in [r_{\mathrm{lo}}^{\mathrm{dec}}, r_{\mathrm{hi}}^{\mathrm{dec}}]$
and
$\bar r^{\mathrm{cov}} \ge r_{\mathrm{lo}}^{\mathrm{cov}}$.\footnote{No upper
edge is needed for the covert cost: $\bar r^{\mathrm{cov}}$ enters the bound
only as a divisor in \cref{eq:cov}, so the worst case sits at its lower edge. Upper limits are uninformative as expensive covert work only hurts the
adversary.} These band edges,
\begin{equation}
	\bigl(r_{\mathrm{hi}}^{\mathrm{dec}},\;r_{\mathrm{lo}}^{\mathrm{dec}},\;
	r_{\mathrm{lo}}^{\mathrm{cov}}\bigr),
	\label{eq:triple}
\end{equation}
are the whole parameter vector of this paper: from here on, all that changes
is which setting of the triple a verifier can justify.

The verifier raises no alarm so long as the measured total is explainable by
honest declared work at some admissible rate and overhead. The most an
honest run could ever draw is the declared work at the top of its band, on
top of the highest overhead,
\begin{equation}
	E \;\le\; r_{\mathrm{hi}}^{\mathrm{dec}}\,C_{\mathrm{dec}} \;+\; E_{0,\mathrm{hi}}.
	\label{eq:noalarm}
\end{equation}
The test is one-sided because covert work can only add energy: only a
total that runs high is suspicious.
Substituting the budget \cref{eq:cheaterbudget} into the no-alarm condition
\cref{eq:noalarm} and rearranging,
\begin{equation}
	C_{\mathrm{cov}}
	\;\le\; \frac{\bigl(r_{\mathrm{hi}}^{\mathrm{dec}} - \bar r^{\mathrm{dec}}\bigr)\,C_{\mathrm{dec}}
	          \;+\; \bigl(E_{0,\mathrm{hi}} - E_0\bigr)}{\bar r^{\mathrm{cov}}}.
	\label{eq:cov}
\end{equation}
A cheater now pushes every free choice on the right-hand side to its limit:
\begin{itemize}
	\item \emph{Run the declared work cheaply,}
	$\bar r^{\mathrm{dec}} \to r_{\mathrm{lo}}^{\mathrm{dec}}$. The verifier
	cannot tell an expensive declared run from a cheap one, so it must allow
	the draw all the way up to $r_{\mathrm{hi}}^{\mathrm{dec}}\,C_{\mathrm{dec}}$;
	the gap
	$(r_{\mathrm{hi}}^{\mathrm{dec}} - r_{\mathrm{lo}}^{\mathrm{dec}})\,C_{\mathrm{dec}}$
	is free energy the prover spends covertly while the total still reads as
	honest work.
	\item \emph{Idle the overhead,} $E_0 \to E_{0,\mathrm{lo}}$, freeing the
	full band width $\Delta E_0 = E_{0,\mathrm{hi}} - E_{0,\mathrm{lo}}$.
	\item \emph{Run the covert work cheaply,}
	$\bar r^{\mathrm{cov}} \to r_{\mathrm{lo}}^{\mathrm{cov}}$, so each hidden
	operation costs as little as the hardware allows.
\end{itemize}
At those values \cref{eq:cov} attains its maximum, the largest covert load
the energy budget admits,
\begin{equation}
	C_{\mathrm{cov}}^{\max,\,\mathrm{en}} \;=\;
	\frac{\bigl(r_{\mathrm{hi}}^{\mathrm{dec}} - r_{\mathrm{lo}}^{\mathrm{dec}}\bigr)\,C_{\mathrm{dec}}
	      \;+\; \Delta E_0}{r_{\mathrm{lo}}^{\mathrm{cov}}}.
	\label{eq:covmax}
\end{equation}
Divide by $C_{\max}$ and write the declared hardware utilisation
$h = C_{\mathrm{dec}}/C_{\max}$ for the fraction of the machine's
declared-format capacity the declared work occupies. The result is a covert load in the
normalised units of \cref{eq:betadef}: the relaxed bound if energy
were the only constraint. We write it $\beta_{\mathrm{en}}$, the
\emph{energy-only width},
\begin{equation}
\;\beta_{\mathrm{en}}(h) \;=\; h\,
		\frac{r_{\mathrm{hi}}^{\mathrm{dec}} - r_{\mathrm{lo}}^{\mathrm{dec}}}{r_{\mathrm{lo}}^{\mathrm{cov}}}
		\;+\; \frac{\Delta E_0}{r_{\mathrm{lo}}^{\mathrm{cov}}\,C_{\max}}\,.\;
	\label{eq:betaen}
\end{equation}
Pedagogically, \cref{eq:betaen} describes (how much honest work there is) $\times$ (how uncertain its
per-operation cost is), divided by (how cheaply covert work can run), plus a
smaller slack from the overhead band. The first term grows with the
declared-band width and with the declared utilisation: more legitimate work
offers more cover. The second term is the overhead's contribution: joules
the prover frees by idling overhead that the verifier must still allow for.

Note the asymmetry between the numerator and denominator of \cref{eq:betaen}.
Declarations and re-execution can narrow the declared band width
$r_{\mathrm{hi}}^{\mathrm{dec}} - r_{\mathrm{lo}}^{\mathrm{dec}}$ in the
numerator, but no declaration reaches the covert floor
$r_{\mathrm{lo}}^{\mathrm{cov}}$ in the denominator; only hardware-wide
observables (clock, voltage) can raise $r_{\mathrm{lo}}^{\mathrm{cov}}$. The
overhead width $\Delta E_0$ is a second numerator, and no declaration reaches
it either: it is freedom the prover exercises inside the window rather than
ignorance about a quantity fixed during it, so characterising it off-site
measures it without shrinking it. It yields only to an on-site observable of
the idle state, or to a restriction on which overhead states may accompany
declared work (\cref{sec:decreasing-beta}).

\subsection{The capacity clip}
\label{subsec:capclip}

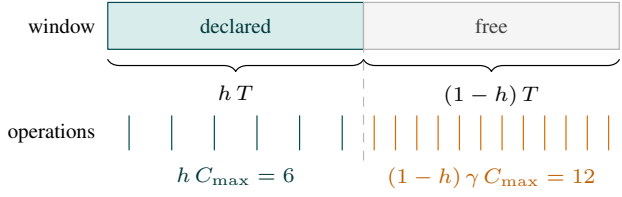
\begin{figure}[t]
	\centering
	\resizebox{\columnwidth}{!}{%
	\begin{tikzpicture}[line join=round, line cap=round]

		\fill[teal!15] (0,1.9) rectangle (3.0,2.45);
		\draw[teal!60!black] (0,1.9) rectangle (3.0,2.45);
		\node[teal!45!black, inner sep=1pt] at (1.5,2.175) {\scriptsize declared};
		\fill[gray!8] (3.0,1.9) rectangle (6.0,2.45);
		\draw[gray!60] (3.0,1.9) rectangle (6.0,2.45);
		\node[gray!50!black, inner sep=1pt] at (4.5,2.175) {\scriptsize free};
		\node[anchor=east, inner sep=4pt] at (0,2.175) {\scriptsize window};

		\draw[decorate, decoration={brace, amplitude=4pt, mirror}]
			(0,1.78) -- (3.0,1.78) node[midway, below=5pt] {\scriptsize $h\,T$};
		\draw[decorate, decoration={brace, amplitude=4pt, mirror}]
			(3.0,1.78) -- (6.0,1.78) node[midway, below=5pt]
			{\scriptsize $(1-h)\,T$};

		\foreach \x in {0.25,0.75,...,2.75}
			\draw[teal!60!black] (\x,0.72) -- (\x,1.12);
		\foreach \x in {3.125,3.375,...,5.875}
			\draw[orange!85!black] (\x,0.72) -- (\x,1.12);
		\draw[gray!55, dashed] (3.0,0.60) -- (3.0,1.9);

		\node[teal!45!black, anchor=north, inner sep=4pt] at (1.5,0.66)
			{\scriptsize $h\,C_{\max} = 6$};
		\node[orange!75!black, anchor=north, inner sep=4pt] at (4.5,0.66)
			{\scriptsize $(1-h)\,\gamma\,C_{\max} = 12$};
		\node[anchor=east, inner sep=4pt] at (0,0.92) {\scriptsize operations};

	\end{tikzpicture}%
	}
	\caption{The capacity clip. Declared work takes $h\,T$ of the window and
		the covert block gets the rest (top); each block executes operations at
		its own format's rate (bottom), drawn at $h = 1/2$ and $\gamma = 2$
		(int8 covert work against declared fp16). The free half-window hides a
		full $C_{\max} = 12$ worth of covert work.}
	\label{fig:timeclip}
\end{figure}

The second constraint is capacity: covert work must also fit on the
machine. The declared and covert streams multiplex one device across a single window of
length $T$, so the covert block runs only in the time the declared block
leaves free. What turns free time into an operation count is a rate. The realised rate
$F(t)$ is the prover's to choose, so a worst-case bound must price the time
at the one rate the verifier does know: the published peak $F_{\max}$. But
the peak is not just a property of the hardware; it is also a property of
the executing format. An A100's dense peaks run from
$19.5$\,Top/s at fp32 on the CUDA cores through $156$ at tf32 and $312$ at
fp16/bf16 to $624$\,Top/s at int8 on the tensor cores \citep{nvidia2020a100}.
We define the capacity ratio of the covert format against the declared one,
\begin{equation}
	\gamma \;=\; \frac{F_{\max}^{\mathrm{cov}}}{F_{\max}^{\mathrm{dec}}}
	\label{eq:capratio}
\end{equation}
($\gamma = 2$ for int8 covert work against fp16 declared work on an A100). The declared work of $C_{\mathrm{dec}} = h\,C_{\max}$ operations cannot be executed
in less than $C_{\mathrm{dec}}/F_{\max}^{\mathrm{dec}} = h\,T$ of the window,
so at most $(1-h)\,T$ is left free (\cref{fig:timeclip}).
Run at the covert format's peak, that free
time is worth $F_{\max}^{\mathrm{cov}}\,(1-h)\,T$ covert operations. Dividing by
$C_{\max} = F_{\max}^{\mathrm{dec}}\,T$ gives
\begin{equation}
	C_{\mathrm{cov}}/C_{\max} \;\le\; (1-h)\,\gamma .
	\label{eq:capclip}
\end{equation}
When $\gamma > 1$ the covert format executes more operations per unit of
machine-time than the declared one, so $C_{\mathrm{dec}} + C_{\mathrm{cov}}$
may exceed $C_{\max}$: the normaliser is the machine's capacity priced in
declared-format operations, not a ceiling on how many operations can
run.

\subsection{The relaxed bound}
\label{subsec:betafloor}

The covert load must fit under both relaxed caps at once, so the bound is the
smaller of the two,
\begin{equation}
	\;\beta(h) \;=\;
	\min\bigl[\,\beta_{\mathrm{en}}(h),\; (1-h)\,\gamma\,\bigr].\;
	\label{eq:betafloor}
\end{equation}

\begin{figure}[t]
	\centering
	\resizebox{\columnwidth}{!}{%
	\begin{tikzpicture}[>=Stealth, line join=round, line cap=round, yscale=1.2]

		\draw[->] (0,0) -- (5.6,0) node[right, inner sep=2pt] {\scriptsize $h$};
		\draw[->] (0,0) -- (0,2.6) node[above, inner sep=2pt] {\scriptsize $\beta$};
		\node[anchor=north east, inner sep=3pt] at (0,0) {\scriptsize $0$};
		\draw (5.0,0) -- (5.0,-0.08)
			node[anchor=north, inner sep=3pt] {\scriptsize $1$};

		\draw[orange!85!black, dashed] (0,0.15) -- (5.0,2.25);
		\node[orange!75!black, anchor=south east, inner sep=2pt] at (5.0,2.25)
			{\scriptsize $\beta_{\mathrm{en}}(h) = s\,h + b$};

		\draw[gray!60!black, dotted, thick] (0,1.5) -- (5.0,0);
		\node[gray!55!black, anchor=south west, inner sep=1pt] at (0.80,1.28)
			{\scriptsize $(1-h)\,\gamma$};

		\draw[blue!55!black, line width=1.1pt]
			(0,0.15) -- (1.875,0.9375) -- (5.0,0);
		\fill[blue!55!black] (1.875,0.9375) circle (1.6pt);

		\draw[gray!55, dashed, thin] (1.875,0.9375) -- (1.875,0);
		\draw[gray!55, dashed, thin] (1.875,0.9375) -- (0,0.9375);
		\node[anchor=north, inner sep=3pt] at (1.875,0) {\scriptsize $h^{*}$};
		\node[anchor=east, inner sep=3pt] at (0,0.9375) {\scriptsize $\beta^{*}$};

		\node[anchor=east, inner sep=3pt] at (0,1.5) {\scriptsize $\gamma$};
		\node[anchor=east, inner sep=3pt] at (0,0.15) {\scriptsize $b$};

		\node[gray!55!black, anchor=north, inner sep=2pt] at (0.95,-0.34)
			{\scriptsize energy binds};
		\node[gray!55!black, anchor=north, inner sep=2pt] at (3.55,-0.34)
			{\scriptsize capacity binds};

	\end{tikzpicture}%
	}
	\caption{The shape of the relaxed bound. The energy-only width
		$\beta_{\mathrm{en}}(h) = s\,h + b$ rises with declared utilisation---more
		honest work offers more cover (dashed)---while the capacity clip
		$(1-h)\,\gamma$ falls, since more honest work leaves less machine free
		(dotted). The bound \cref{eq:betafloor} is the lower of the two (solid):
		a tent peaking at the crossing $(h^{*}, \beta^{*})$ of
		\cref{eq:betaworst}, with energy binding to its left, capacity to its
		right, and $\beta(1) = 0$ because a fully occupied machine leaves
		nowhere to hide.}
	\label{fig:betashape}
\end{figure}
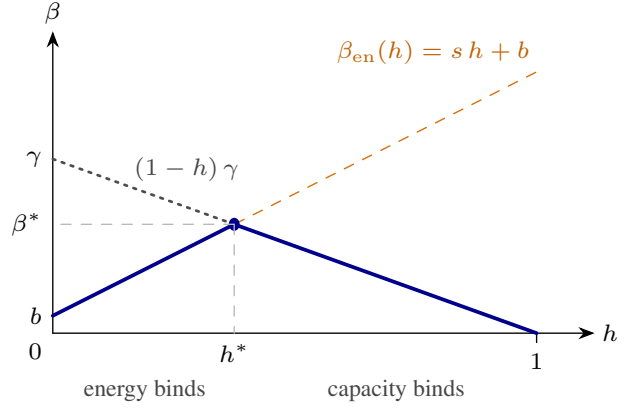

The two caps pull in opposite directions (\cref{fig:betashape}). Energy
rises with $h$---more declared work offers more cover---while capacity
falls---more
declared work leaves less machine to hide on. At full declared utilisation
the machine is spoken for and $\beta(1) = 0$; at idle there
is capacity but almost no cover. The declared utilisation is the prover's to
set, so the guarantee a verifier can quote is the worst case over it,
$\beta^{*} \equiv \max_h \beta(h)$. The corresponding physical quantity is
$\beta_{\mathrm{ID}}^{*} \equiv \max_h \beta_{\mathrm{ID}}(h)$, and the
relaxation guarantees $\beta_{\mathrm{ID}}^{*} \leq \beta^{*}$.
The paper tracks the computable bound $\beta^{*}$. Its maximum sits at the crossing,
\begin{equation}
	h^{*} = \frac{\gamma - b}{s + \gamma}, \qquad
	\beta^{*} = \gamma\,\frac{s + b}{s + \gamma},
	\label{eq:betaworst}
\end{equation}
writing $\beta_{\mathrm{en}}(h) = s\,h + b$ for its slope and intercept. The crossing form
holds when $b < \gamma$; if the overhead term alone exceeds the clip
($b \ge \gamma$) the clip binds everywhere and the maximum sits at the
boundary, $(h^{*}, \beta^{*}) = (0, \gamma)$. Each side of the
minimum is the binding constraint in its own regime: energy binds for
$h < h^{*}$ (the trace itself caps the covert load), capacity binds for
$h > h^{*}$, and both bind at the crossing. Where the crossing sits is an
empirical question about $s$.

The relaxation is exact only if one schedule can attain the three selected
edges together---cheap declared work, low overhead, and cheap covert work.
No such joint attainment is assumed, so throughout
$\beta_{\mathrm{ID}}(h) \leq \beta(h)$. \Cref{app:boundkind} discusses this
joint-attainment assumption and brackets the remaining gap with a constructed schedule.

\section{How big is \texorpdfstring{$\beta$}{beta}?}
\label{sec:howbig}

We now estimate the typical value of $\beta$, \cref{eq:betafloor}, for real hardware. We first ask what physically drives each entry of the triple \cref{eq:triple}
(\cref{subsec:factors}), then measure them directly on an NVIDIA A100, where
each experiment is built to pin one entry
(\cref{sec:empirical-measurements}). We focus on the $\beta_{\rm en}(h)$ component, \cref{eq:betaen}: it carries all
the physics, whereas the clip \cref{eq:capclip} is a ratio of published peaks
with nothing to measure. The capacity clip \cref{eq:capclip}
returns at the end, when the measured bands are assembled.

\subsection{The two terms}
\label{subsec:factors}

The energy-only width $\beta_{\rm en}(h)$ is a sum of two terms: an ambiguity term
$s\,h$ with slope
$s = (r_{\mathrm{hi}}^{\mathrm{dec}} - r_{\mathrm{lo}}^{\mathrm{dec}})/r_{\mathrm{lo}}^{\mathrm{cov}}$,
set by the spread of the declared exchange rate against the price of a covert
operation, and an overhead term
$b = \Delta E_0/(r_{\mathrm{lo}}^{\mathrm{cov}}\,C_{\max})$, set by the width of
the overhead band against the same price. We take each term in turn.

\paragraph{The exchange-rate term.}
\Cref{subsec:forward} wrote the exchange rate as
$r = \alpha\,(C/N_{\mathrm{op}})\,V^2$. Here we split the capacitance term in
two, since the numeric format and the operand locality move independently:
\begin{equation}
	r \;\approx\;
	\underbrace{\kappa}_{\text{precision}} \,\cdot\,
	\underbrace{\alpha}_{\text{operands}} \,\cdot\,
	\underbrace{C_{\mathrm{eff}}}_{\text{locality, kernel}} \,\cdot\,
	\underbrace{V^2}_{\text{DVFS}}.
	\label{eq:cmos}
\end{equation}
\Cref{app:cmos} derives this factorisation. Each factor is set by a different physical scale of the
machine (\cref{fig:exchange-scales}). The numeric-format scale $\kappa$ decreases
with datapath width at each precision step (fp32 $\to$ int8), tracking the
pJ-scale energy hierarchy of arithmetic \citep{horowitz2014computing}; the
activity factor $\alpha$ is the data-dependent fraction of transistors that
toggle for the given operands; the effective capacitance $C_{\mathrm{eff}}$
grows as operands become less local, because more of the memory
hierarchy (e.g. storage arrays and peripheral logic, buffers and interfaces,
interconnect) switches per operation
\citep{horowitz2014computing,chatterjee2017architecting}; and $V^2$ is the
capacitor-charging energy of the supply voltage, which DVFS slides along the
voltage--frequency curve \citep{pallipadi2006ondemand}.

With the power meter as the only channel, every factor of \cref{eq:cmos} is
conceded at once. If each ranges over an interval the verifier cannot narrow,
and the factors can be extreme together, the declared band's edges are the
products of the factor edges,
\begin{equation}
	r_{\mathrm{hi}}^{\mathrm{dec}} =
	\kappa_{\mathrm{hi}}\,\alpha_{\mathrm{hi}}\,C_{\mathrm{hi}}\,V_{\mathrm{hi}}^2,
	\qquad
	r_{\mathrm{lo}}^{\mathrm{dec}} =
	\kappa_{\mathrm{lo}}\,\alpha_{\mathrm{lo}}\,C_{\mathrm{lo}}\,V_{\mathrm{lo}}^2,
	\label{eq:edgefactor}
\end{equation}
writing $C$ for $C_{\mathrm{eff}}$ to keep the subscripts readable. Dividing,
the band's ratio is the product of per-factor widths
\begin{equation}
	\frac{r_{\mathrm{hi}}^{\mathrm{dec}}}{r_{\mathrm{lo}}^{\mathrm{dec}}}
	\;=\; w_\kappa\,w_\alpha\,w_C\,w_V,
	\qquad w_x \equiv \frac{x_{\mathrm{hi}}}{x_{\mathrm{lo}}}\,,
	\qquad w_V \equiv \frac{V_{\mathrm{hi}}^2}{V_{\mathrm{lo}}^2}\,,
	\label{eq:widthfactor}
\end{equation}
each $w_x \ge 1$. Taking every factor at its widest together makes this an upper bound, for
the same reason the caveat closing \cref{subsec:betafloor} gave for $\beta$
itself; \cref{app:boundkind} treats both. Writing
$W \equiv w_\kappa w_\alpha w_C w_V$ we can express the slope as 
\begin{equation}
	s \;=\;
	\frac{r_{\mathrm{hi}}^{\mathrm{dec}} - r_{\mathrm{lo}}^{\mathrm{dec}}}
	     {r_{\mathrm{lo}}^{\mathrm{cov}}}
	\;=\; \frac{r_{\mathrm{lo}}^{\mathrm{dec}}}{r_{\mathrm{lo}}^{\mathrm{cov}}}
	      \,\bigl(W - 1\bigr).
	\label{eq:slopebridge}
\end{equation}
Note  we never evaluate $\beta$ from the factorisation; every number we report comes
from measured band edges directly.

\begin{figure*}[t]
	\centering
	\resizebox{\textwidth}{!}{%
	\begin{tikzpicture}[
		>=Stealth, line join=round, line cap=round,
		panel/.style={draw=black!16, fill=black!1, rounded corners=3pt,
			line width=0.6pt},
		badge/.style={rounded corners=2pt, inner sep=2.5pt,
			font=\footnotesize},
		note/.style={font=\scriptsize, align=center},
		title/.style={font=\footnotesize},
		tiny/.style={font=\tiny},
		mos/.style={font=\tiny, draw=orange!85!black, fill=orange!7,
			rounded corners=1pt, minimum width=0.62cm, minimum height=0.33cm,
			inner sep=1pt},
		logic/.style={font=\scriptsize, draw=blue!55!black, fill=blue!5,
			circle, minimum size=0.42cm, inner sep=0pt},
		fma/.style={font=\tiny, draw=blue!55!black, fill=blue!8,
			rounded corners=1pt, minimum height=0.4cm, inner sep=1.5pt},
		mem/.style={font=\tiny, draw=teal!60!black, fill=teal!4,
			rounded corners=1pt, minimum height=0.4cm, inner sep=1.3pt}]

		\begin{scope}[shift={(0,0)}]
			\draw[panel] (0,0.1) rectangle (3.55,4.25);
			\node[title] at (1.775,3.98) {(1) CMOS switch};
			\draw[black!12] (0.18,3.72) -- (3.37,3.72);

			\node[mos] (pmos) at (1.38,3.35) {pMOS};
			\node[mos] (nmos) at (1.38,2.63) {nMOS};
			\draw[orange!85!black, line width=0.65pt]
				(1.38,3.62) -- (pmos)
				(pmos) -- (1.38,2.99) -- (nmos) -- (1.38,2.22);
			\node[tiny, black, anchor=east] at (1.31,3.60) {\(V\)};
			\draw[orange!85!black] (1.23,2.18) -- (1.53,2.18)
				(1.28,2.10) -- (1.48,2.10) (1.33,2.02) -- (1.43,2.02);
			\fill[orange!85!black] (1.38,2.99) circle (1.2pt);
			\draw[->, orange!85!black] (0.35,2.99) -- (0.95,2.99)
				node[tiny, black, midway, above] {input};
			\draw[orange!85!black] (0.95,2.99) -- (0.95,3.35) -- (pmos.west)
				(0.95,2.99) -- (0.95,2.63) -- (nmos.west);
			\draw[orange!85!black] (1.38,2.99) -- (2.42,2.99) -- (2.42,2.82);
			\draw[orange!85!black, line width=0.7pt] (2.27,2.82) -- (2.57,2.82);
			\draw[orange!85!black, line width=0.7pt] (2.27,2.69) -- (2.57,2.69);
			\draw[orange!85!black] (2.42,2.69) -- (2.42,2.22);
			\draw[orange!85!black] (2.27,2.18) -- (2.57,2.18)
				(2.32,2.10) -- (2.52,2.10) (2.37,2.02) -- (2.47,2.02);
			\node[tiny, anchor=west] at (2.62,2.76) {load \(c_i\)};
			\node[tiny, anchor=south] at (2.02,3.02) {output};

			\foreach \x/\c in {0.82/gray!28,1.20/orange!85!black,
				1.58/orange!85!black,1.96/gray!28,2.34/gray!28,2.72/orange!85!black}
				\fill[\c] (\x-0.12,1.52) rectangle (\x+0.12,1.76);
			\node[tiny, text=orange!85!black] at (1.775,1.31)
				{only changing cells charge their loads};
			\node[badge, fill=orange!85!black, text=white] at (1.775,0.88)
				{\(\alpha\)};
			\node[note, text=orange!85!black] at (1.775,0.43)
				{operand-dependent\\toggling fraction};
		\end{scope}

		\draw[->, thick] (3.63,2.36) -- (4.00,2.36);

		\begin{scope}[shift={(4.08,0)}]
			\draw[panel] (0,0.1) rectangle (3.55,4.25);
			\node[title] at (1.775,3.98) {(2) FMA datapath};
			\draw[black!12] (0.18,3.72) -- (3.37,3.72);

			\node[tiny] (ain) at (0.40,3.35) {\(a\)};
			\node[tiny] (bin) at (0.40,2.99) {\(b\)};
			\node[logic] (mul) at (1.30,3.17) {\(\times\)};
			\node[logic] (add) at (2.34,3.17) {\(+\)};
			\node[tiny] (cin) at (1.70,2.75) {\(c\)};
			\draw[->, blue!55!black] (ain) -- (mul);
			\draw[->, blue!55!black] (bin) -- (mul);
			\draw[->, blue!55!black] (mul) -- (add);
			\draw[->, blue!55!black] (cin) -- (add);
			\draw[->, blue!55!black] (add) -- (3.12,3.17)
				node[tiny, black, right] {\(y\)};
			\foreach \x in {1.12,1.29,1.46,2.16,2.33,2.50}
				\fill[orange!85!black]
					(\x-0.035,3.405) rectangle (\x+0.035,3.475);
			\node[tiny, text=orange!85!black] at (1.82,3.62) {networks of CMOS cells};

			\node[tiny, anchor=east] at (0.58,2.40) {fp32};
			\draw[blue!55!black, fill=blue!13] (0.70,2.26) rectangle (3.05,2.54);
			\node[tiny] at (1.875,2.40) {one 32-bit word};
			\node[tiny, anchor=east] at (0.58,1.95) {fp16};
			\foreach \x in {0.70,1.89}
				\draw[blue!55!black, fill=blue!9] (\x,1.81) rectangle (\x+1.16,2.09);
			\node[tiny] at (1.28,1.95) {16};
			\node[tiny] at (2.47,1.95) {16};
			\node[tiny, anchor=east] at (0.58,1.50) {int8};
			\foreach \x in {0.70,1.30,1.90,2.50}
				\draw[blue!55!black, fill=blue!6] (\x,1.36) rectangle (\x+0.55,1.64);
			\foreach \x in {0.975,1.575,2.175,2.775}
				\node[tiny] at (\x,1.50) {8};
			\node[badge, fill=blue!55!black, text=white] at (1.775,0.88)
				{\(\kappa\)};
			\node[note, text=blue!55!black] at (1.775,0.43)
				{format sets gates and\\results per datapath};
		\end{scope}

		\draw[->, thick] (7.71,2.36) -- (8.08,2.36);

		\begin{scope}[shift={(8.16,0)}]
			\draw[panel] (0,0.1) rectangle (3.55,4.25);
			\node[title] at (1.775,3.98) {(3) memory hierarchy};
			\draw[black!12] (0.18,3.72) -- (3.37,3.72);
			\node[mem, minimum width=0.55cm] (dram) at (0.45,2.87) {DRAM};
			\node[mem, minimum width=0.58cm] (cache) at (1.32,2.87) {cache};
			\node[mem, minimum width=0.48cm] (reg) at (2.18,2.87) {regs};
			\node[fma, minimum width=0.52cm, line width=0.8pt]
				(memfma) at (3.02,2.87) {FMA};
			\draw[->, teal!60!black, line width=0.7pt] (dram) -- (cache);
			\draw[->, teal!60!black, line width=0.7pt] (cache) -- (reg);
			\draw[->, teal!60!black, line width=0.7pt] (reg) -- (memfma);
			\node[tiny, text=teal!60!black] at (1.73,3.34)
				{read \(a,b,c\)};
			\draw[<-, teal!60!black, dashed, line width=0.55pt]
				(dram.south) .. controls (1.25,2.12) and (2.60,2.12) .. (memfma.south);
			\node[tiny, text=teal!60!black, fill=black!1, inner sep=1pt]
				at (1.75,2.18) {write result \(y\)};
			\draw[decorate, decoration={brace, amplitude=4pt, mirror}, teal!60!black]
				(0.22,1.82) -- (3.27,1.82);
			\node[tiny, text=teal!60!black, align=center] at (1.745,1.48)
				{more levels crossed \(\Rightarrow\) more circuits\\and capacitance switched};
			\node[badge, fill=teal!60!black, text=white] at (1.775,0.88)
				{\(C_{\mathrm{eff}}\)};
			\node[note, text=teal!60!black] at (1.775,0.43)
				{locality and kernel set\\capacitance per result};
		\end{scope}

		\draw[->, thick] (11.79,2.36) -- (12.16,2.36);

		\begin{scope}[shift={(12.24,0)}]
			\draw[panel] (0,0.1) rectangle (3.55,4.25);
			\node[title] at (1.775,3.98) {(4) voltage domain};
			\draw[black!12] (0.18,3.72) -- (3.37,3.72);
			\draw[violet!70!black, fill=violet!2, rounded corners=2.5pt]
				(0.48,1.72) rectangle (3.10,3.60);
			\draw[violet!70!black, line width=1.05pt] (0.75,3.35) -- (2.84,3.35);
			\node[tiny, text=violet!70!black, fill=violet!2, inner sep=1pt,
				anchor=north] at (1.795,3.31) {supply rail};
			\draw[violet!70!black, line width=0.8pt] (0.75,1.95) -- (2.84,1.95);
			\node[tiny, text=violet!70!black, fill=violet!2, inner sep=1pt,
				anchor=north] at (1.795,1.92) {ground rail};
			\node[mem] (diemem) at (1.12,2.55) {cache/regs};
			\node[fma] (diefma) at (2.46,2.55) {FMA};
			\draw[->, teal!60!black] (diemem) -- (diefma);
			\draw[->, violet!70!black] (1.12,3.35) -- (diemem);
			\draw[->, violet!70!black] (2.46,3.35) -- (diefma);
			\draw[violet!70!black] (diemem.south) -- (1.12,1.95);
			\draw[violet!70!black] (diefma.south) -- (2.46,1.95);
			\draw[<->, violet!70!black, line width=0.8pt]
				(0.38,1.95) -- (0.38,3.35);
			\node[tiny, text=violet!70!black, anchor=east] at (0.34,2.65)
				{\(V\)};
			\node[tiny, text=violet!70!black] at (1.79,1.52)
				{DVFS changes every switching cost};
			\node[badge, fill=violet!70!black, text=white] at (1.775,0.88)
				{\(V^2\)};
			\node[note, text=violet!70!black] at (1.775,0.43)
				{energy to charge each\\switched capacitance};
		\end{scope}

		\draw[decorate, decoration={brace, amplitude=5pt, mirror}, black!45]
			(4.25,-0.05) -- (11.55,-0.05);
		\node[font=\scriptsize] at (7.90,-0.50)
			{format and locality set capacitance per operation};

		\draw[black!20] (0,-0.88) -- (15.79,-0.88);
		\node[font=\normalsize] at (7.895,-1.47)
			{\(r \;=\;
			{\color{orange!85!black}\alpha}\,\cdot\,
			{\color{blue!55!black}\kappa}\,\cdot\,
			{\color{teal!60!black}C_{\mathrm{eff}}}\,\cdot\,
			{\color{violet!70!black}V^2}\)};
		\node[font=\scriptsize, text=gray!60!black] at (7.895,-1.96)
			{\(f\) cancels from the marginal rate; leakage and cooling leave as
			the overhead \(P_0\), not \(r\)};

	\end{tikzpicture}%
	}
	\caption{\textbf{The exchange rate, from transistors to the powered die.}
		Each physical scale contributes one factor to the per-operation energy of
		\cref{eq:cmos}. (1)~A complementary pMOS--nMOS inverter charges or discharges
		its effective load \(c_i\) when its input changes; an operand determines
		what fraction \(\alpha\) of all such cells toggle. (2)~Networks of these
		cells implement the multiply and add of an FMA. The numeric format selects
		the word width (and supported packing), changing the logic engaged per
		arithmetic result and hence \(\kappa\). (3)~The FMA reads operands through
		the register--cache--DRAM hierarchy and writes its result back. Traversing
		more of that hierarchy switches additional array and peripheral circuits,
		buffers and interfaces, and interconnect; their combined dynamic cost is
		represented by \(C_{\mathrm{eff}}\). (4)~Within a voltage domain, the supply places
		\(V\) across the arithmetic and interconnect loads, so DVFS contributes
		\(V^2\) per unit capacitance. The
		clock \(f\) cancels from the marginal rate (\cref{eq:rdyn}) and the static
		draw---leakage, cooling---leaves as the overhead \(P_0\).
		Multiplying the factors recovers \cref{eq:cmos}.}
	\label{fig:exchange-scales}
\end{figure*}

\paragraph{The overhead term.}
The overhead $P_0$ admits no comparably
clean factorisation. It is a
heterogeneous mix of static leakage, cooling control loops, firmware idle states,
and platform I/O, so no single relation fixes the band width $\Delta E_0$ from
first principles.

\subsection{Measurement on an NVIDIA A100}
\label{sec:empirical-measurements}
In order to measure $\beta$ we require estimates of the three entries of the triple $	\bigl(r_{\mathrm{hi}}^{\mathrm{dec}},\;r_{\mathrm{lo}}^{\mathrm{dec}},\;
	r_{\mathrm{lo}}^{\mathrm{cov}}\bigr)$, 
\cref{eq:triple}, and the overhead width $\Delta E_0$. Each experiment below
exists to pin one of them. Experiment~1 supplies both declared edges
$(r_{\mathrm{hi}}^{\mathrm{dec}}, r_{\mathrm{lo}}^{\mathrm{dec}})$,
Experiment~2 supplies $\Delta E_0$, and Experiment~3 supplies the covert floor
$r_{\mathrm{lo}}^{\mathrm{cov}}$. Results are summarised in \cref{tab:inputs}. Two further
experiments are used later: Experiment~4 measures int8 transformer blocks
(\cref{sec:decreasing-beta}) and Experiment~5 sweeps the operand values at
fixed precision (\cref{app:operands}).

\paragraph{Setup and provenance.}
The experiments below ran sequentially on one NVIDIA A100-SXM4-80GB in a single
Slurm allocation. The platform denied clock locking and power capping, so DVFS is a
recorded axis rather than a controlled one: each window logs the
streaming-multiprocessor (SM) clock selected by the driver's frequency
governor, and the bands are stratified into clock strata. That also puts $V^2$ beyond measurement here, so of the four factors of
\cref{eq:cmos} we measure three and carry $w_V \approx 1.5$ from device physics
throughout; this is discussed as a limitation in \cref{sec:discussion}.\footnote{Supply
voltages on modern accelerators sit near $1$\,V \citep{horowitz2014computing}
and DVFS moves them by roughly $\pm 10\%$; energy per operation scales as the
ratio squared, $(1.1/0.9)^2 \approx 1.5$.}
Energy is read from the on-die NVML total-energy counter; each
measured cell fills a ${\sim}1.5$\,s window with back-to-back kernels on
pre-allocated buffers and keeps the per-cell median over repeated
windows. This is a non-adversarial, device-level calibration: the
telemetry is trusted as a laboratory instrument. One GPU
suffices to estimate the bands and the order of magnitude of $\beta$;
extending the calibration to off-chip meters and multi-device systems is
natural future work, taken up in \cref{sec:discussion}. \Cref{app:bench} gives the full protocol, the meter
tolerance, and the provenance manifest, including which card measured
which band.

\begin{table}[t]
	\centering
	\small
	\begingroup
	\fontsize{8}{9.5}\selectfont
	\setlength{\tabcolsep}{4pt}
	\begin{tabular}{@{}lllr@{}}
		\toprule
		quantity & what it is & source & value \\
		\midrule
		\multicolumn{4}{@{}l}{\emph{(a) the factor widths of \cref{eq:widthfactor}}} \\
		$w_\kappa$ & precision & Exp.~1, cond.\ & $13.5\times$ \\
		$w_\alpha$ & operands & Exp.~1, cond.\ & $1.88\times$ \\
		$w_C$ & locality & Exp.~1, cond.\ & $1.04\times$ \\
		$w_V$ & clock (DVFS) & \emph{not swept; assumed} & ${\approx}1.5\times$ \\
		\addlinespace[1pt]
		$w_\kappa w_\alpha w_C$ & product & --- & $26.5\times$ \\
		\midrule
		\multicolumn{4}{@{}l}{\emph{(b) the edges the bound reads, nothing pinned}} \\
		$r_{\mathrm{hi}}^{\mathrm{dec}}$ & fp32, expensive operands & Exp.~1 & $19.5$ \\
		$r_{\mathrm{lo}}^{\mathrm{dec}}$ & int8, zeros & Exp.~1 & $0.786$ \\
		$r_{\mathrm{lo}}^{\mathrm{cov}}$ & generic covert op & Exp.~3 & $0.51$ \\
		$\Delta P_0$ & overhead band & Exp.~2 & $41.2$\,W \\
		\bottomrule
	\end{tabular}
	\endgroup
	\caption{Summary of results from Experiments 1--3. Values in pJ/op unless marked
	otherwise.
	Block (a) is the physics: the four factors of \cref{eq:cmos} and how wide
	each one runs. ``Cond.'' means the width is measured with the other three
	factors held fixed, so the widths are separable rather than pooled. Their
	product reproduces the measured span to $7\%$, and $w_\alpha w_C = 1.96$
	predicts the fp16 sub-band's measured $2.00$ to $2\%$.
	$w_V$ is the exception and is flagged as such: Exp.~1 does not sweep the
	clock (the governor held it inside a $\pm3\%$ window), so its width is a
	physics estimate, not a measurement.
	Block (b) is what the bound reads: the triple \cref{eq:triple} and the
	overhead width. These are the values a bare power meter concedes, with no
	format pinned on either declared edge and the covert floor priced for
	generic covert work. $C_{\max}$ and $\gamma$ are read from published peaks rather
	than measured here.}
	\label{tab:inputs}
\end{table}

\begin{figure}[t]
	\centering
	\includegraphics[width=\columnwidth]{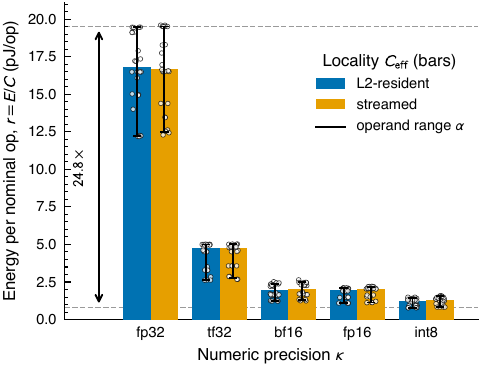}
	\caption{Experiment~1: Total energy quotient $E/C$ for a fixed matmul kernel on an NVIDIA
	A100, swept factorially over precision $\kappa$ (bar groups), operand
	locality $C_{\mathrm{eff}}$ (paired bars), and operand distribution
	$\alpha$ (whiskers); white points are the individual measured windows
	($n=240$ across $60$ cells). Each bar is the median over the operand
	axis and its whisker that axis's full range, so the operand factor (a
	little under $2\times$ inside every format, against locality's couple of
	per cent) is read off the figure rather than pooled into it. Dashed lines
	mark the widest setting's edges, both of them operand extremes: fp32 at its
	most expensive operands, int8 with zeros.}
	\label{fig:exchange-band}
\end{figure}

\paragraph{Experiment 1---the declared band
$(r_{\mathrm{hi}}^{\mathrm{dec}}, r_{\mathrm{lo}}^{\mathrm{dec}})$, and three of
the four factor widths $(w_\kappa, w_\alpha, w_C)$.}
A compute-bound matmul (operation
count $C = 2MNK$ known analytically%
\footnote{The int8 cells retire integer multiply--adds, counted as nominal operations
per the convention of \cref{subsec:forward}.}),
swept factorially over precision ($\kappa$, fp32 down to int8), operand
locality ($C_{\mathrm{eff}}$, operands resident in vs.\ streamed through
the L2 at a byte-identical shape), and operand
distribution ($\alpha$, zeros through adversarial-toggle).

The bound's two declared entries come off that sweep directly. With no format
pinned they are its extremes, the most expensive cell (fp32, adversarial-toggle
operands) against the cheapest (int8, zeros; \cref{fig:exchange-band}),
\begin{equation}
	r_{\mathrm{hi}}^{\mathrm{dec}} = 19.5\;\mathrm{pJ/op}, \qquad
	r_{\mathrm{lo}}^{\mathrm{dec}} = 0.786\;\mathrm{pJ/op}.
	\label{eq:measureddec}
\end{equation}

The sweep is a factorial so the widths of \cref{eq:widthfactor} are also obtained for free.
The $60$ cells are every combination of five
precisions, two localities and six operand distributions. To measure one
factor's width, hold the other two axes fixed and take the ratio of the most to
the least expensive cell as that factor runs over all its levels; the width is
the median of that ratio across every such group of cells. This gives
\begin{equation}
	w_\kappa = 13.5, \qquad w_\alpha = 1.88, \qquad w_C = 1.04,
	\label{eq:measuredwidths}
\end{equation}
with precision following the textbook ordering
int8 $<$ bf16 $\approx$ fp16 $<$ tf32 $<$ fp32. So precision dominates, operand
values are worth a factor near two, and locality moves $r$ by four per cent for
this compute-bound kernel.\footnote{Per access, DRAM costs about a hundred times more than on-chip
memory \citep{horowitz2014computing}; the per-operation effect is diluted by the
kernel's bytes-per-operation, which is why $w_C$ is small here. A memory-bound
kernel would not enjoy that dilution, and $w_C$ is the one width we would expect
to move substantially with the workload.} The clock is the one axis we could not hold fixed, so we check it rather than
group on it. Comparing the median achieved clock across each axis's levels,
the operand and locality levels ran at an identical clock and the precision
levels to within $2.7\%$, so these widths are not clock artefacts. For the same
reason the fourth width is missing: the governor held the clock inside
$1335$--$1410$\,MHz across the whole sweep, a $\pm 3\%$ window, which is what
keeps the three measured widths clean but leaves $w_V$ no range to be measured
over. 

Multiplying the three measured widths gives
$13.5 \times 1.88 \times 1.04 = 26.5$, against the $24.8\times$ span the pooled
cells actually show---so \cref{eq:widthfactor} holds to $7\%$, the residual
being factor interaction. Similarly, among the fp16 cells the precision never varies, so
the spread should be the product of the two remaining widths,
$w_\alpha w_C = 1.96$; those cells span $[1.10,\,2.19]$\,pJ/op, a factor of
$2.00$. \Cref{app:operands} performs a separate fp16-only sweep of the operand axis with a governor-selected clock and corroborates $w_\alpha$ independently, reaching $1.83$ compared to the quoted value here of $1.88$.

\paragraph{Experiment 2---the overhead band $\Delta P_0$ (hence $\Delta E_0$).}
The overhead is what the meter reads with no declared work running. We measure five
variants chosen to bracket the plausible range: before any CUDA context,
context up, context plus operand pool resident, immediately after a sustained
heavy kernel, and after a $60$\,s cooldown. Each variant is measured over five
repeated windows. The band is the lowest and highest of those $25$ windows, not
a range over per-variant medians. The lowest is a cold card before any context,
at $62.9$\,W. The highest is a window taken immediately after the heavy kernel,
at $96.2$\,W. That variant is a transient rather than a steady state: its five
consecutive windows decay from $96.2$ to $76.5$\,W as the card settles. We take its peak
deliberately. An adversary can leave a device in that state, and the meter does
not report which idle state it is reading, so the peak belongs in the envelope.
Reducing each variant to a median instead would give $\Delta P_0 = 23.9$\,W.
The observed range $[62.9,\,96.2]$\,W, stretched outward by the adopted $\pm 5\%$ meter
tolerance\footnote{NVML publishes no accuracy specification for the total-energy
counter \texttt{nvmlDeviceGetTotalEnergyConsumption} on GA100; the documented
$\pm 5\%$ is for the power readings of the earlier Fermi and Kepler generations
\citep{nvidia_nvml}. We adopt it as a working tolerance and apply it outward
only, so it can only widen $\beta$ (\cref{app:bench}). It accounts for
$7.9$\,W of the width below; the remaining $33.3$\,W is genuine spread between
idle states.}, pins the overhead band to
\begin{equation}
	P_0 \in [59.8,\, 101.0]\;\mathrm{W}, \qquad
	\Delta P_0 = 41.2\;\mathrm{W}.
	\label{eq:measuredoverhead}
\end{equation}
Alongside the idle cells, Experiment~2 also sweeps a ladder of ten matmul
shapes spanning a $60$--$90\times$ range of sustained rate, as a consistency
check on the affine forward map \cref{eq:forward}: fitted like for like over
the compute-bound shapes, the ladder's slope lands inside the fp16 exchange
band of Experiment~1. The fitted slope is only ever a diagnostic, never an
input to the bound (\cref{app:estimands}).

\begin{figure}[t]
	\centering
	\includegraphics[width=\columnwidth]{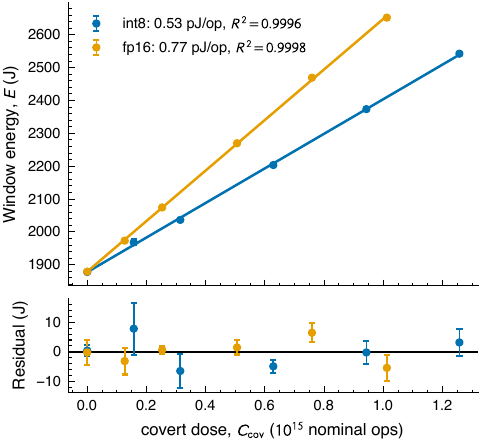}
	\caption{Experiment~3: window energy $E$ vs.\
	covert operation count $C_{\mathrm{cov}}$ for int8 and fp16 covert work,
	added to a
	fixed fp16 declared burst inside a fixed $10$\,s window. \emph{Top:}
	per-cell medians with $\pm1$\,SD error bars over each cell's four
	untrimmed repeats (of five captured; the flagged high-tail repeat is
	excluded),
	against the fit $E = r_{\mathrm{cov}}\,C_{\mathrm{cov}} + E_0$; both
	slopes are linear to $R^2 > 0.999$ ($0.53$\,pJ/op int8, $0.77$\,pJ/op
	fp16) and share an intercept matching the directly measured covert-free
	windows to $0.02\%$. \emph{Bottom:} fit residuals.}
	\label{fig:marginal-dose}
\end{figure}

\paragraph{Experiment 3---the marginal covert cost
$r_{\mathrm{lo}}^{\mathrm{cov}}$.}
We construct a $10$\,s observation window. Inside it we run a fixed fp16
declared burst---a $2048^3$ matmul on declared-typical operands, repeated
enough times to keep the device busy for $30\%$ of the window---then
$N_{\mathrm{cov}}$ covert matmuls of a single format (int8 or fp16, each swept
as its own arm), then an idle pad that
fills out the remainder. The covert operands are zeros, the adversary's
cheapest case. The two formats bracket what matters: int8 is the cheapest
arithmetic the device offers, and fp16 matches the declared work, so covert
operations are indistinguishable from declared ones by format alone. Only
$N_{\mathrm{cov}}$ changes between windows: the declared
burst, the shapes, the operand buffers, and the $10$\,s length are all held
fixed.

To see how this experiment determines $r_{\mathrm{lo}}^{\mathrm{cov}}$, write the window energy out. Let
$e_{\mathrm{cov}}$ be the energy of one covert matmul, $t_{\mathrm{cov}}$ its
duration, $c$ its nominal operation count, $t_{\mathrm{dec}}$ the declared
burst's busy time, and $P_{\mathrm{idle}}$ the power drawn during the pad. Then
\begin{equation}
	E \;=\; E_{\mathrm{dec}}
	\;+\; N_{\mathrm{cov}}\,e_{\mathrm{cov}}
	\;+\; P_{\mathrm{idle}}\bigl(T - t_{\mathrm{dec}}
	- N_{\mathrm{cov}} t_{\mathrm{cov}}\bigr),
	\label{eq:dosewindow}
\end{equation}
which collects into a constant plus a term linear in $N_{\mathrm{cov}}$,
\begin{equation}
	E \;=\; \underbrace{E_{\mathrm{dec}} + P_{\mathrm{idle}}
	\bigl(T - t_{\mathrm{dec}}\bigr)}_{\text{independent of }N_{\mathrm{cov}}}
	\;+\; N_{\mathrm{cov}}
	\underbrace{\bigl(e_{\mathrm{cov}}
	- t_{\mathrm{cov}}P_{\mathrm{idle}}\bigr)}_{\text{the slope}} .
	\label{eq:doseslope}
\end{equation}
The first bracket does not depend on $N_{\mathrm{cov}}$. The covert work
amounts to $C_{\mathrm{cov}} = N_{\mathrm{cov}}\,c$ operations, so
differentiating leaves only the second,
\begin{equation}
	\frac{\partial E}{\partial C_{\mathrm{cov}}}
	\;=\; \frac{e_{\mathrm{cov}} - t_{\mathrm{cov}}P_{\mathrm{idle}}}{c}
	\;\equiv\; r^{\mathrm{cov}} .
	\label{eq:doserate}
\end{equation}

We sweep $N_{\mathrm{cov}}$ over six levels, from no covert work up to a covert
busy fraction of $0.48$, with five repeats per level in randomised order with
drift-reference splices. Regressing window energy on covert nominal operations
then gives the marginal cost directly (\cref{fig:marginal-dose}):
$0.53$\,pJ/op for int8 and $0.77$\,pJ/op for fp16, both linear to
$R^2 > 0.999$. Pricing the same covert kernels the obvious way instead---total
energy divided by total operations---gives $0.26$ (int8) to $0.39$\,pJ/op
(fp16) more. That excess is baseline power charged to the covert operations,
which is what the fixed window differences away. \Cref{app:estimands} states
precisely which estimand each band edge requires, and why the covert edge
must be this marginal cost rather than a quotient. We take the one-sided $95\%$ lower confidence limits,
\begin{equation}
	r_{\mathrm{lo}}^{\mathrm{cov}} = 0.51\;\mathrm{pJ/op}\ \ (\text{int8}),
	\qquad
	0.76\;\mathrm{pJ/op}\ \ (\text{fp16}).
	\label{eq:measuredcov}
\end{equation}
Because the covert cost divides the slack, under-pricing covert work
is the security-safe direction. 

One caveat is that, similar to the other experiments, clock locking is denied on this platform, so the governor is
recorded rather than controlled here too. More covert work leaves a shorter
idle pad, and the window-median clock rises with the load accordingly, from
$1275$\,MHz with no covert work to $1410$\,MHz at the heaviest. However, what changes is
the time spent at each clock frequency, not the operating point the kernels run at:
every measured window, in both arms, records the same $1275$\,MHz minimum and
$1410$\,MHz maximum. We would rather have locked the clock and compared every cell at one
fixed setting. Two checks stand in for that. A clock inflating the cost of
later operations would bend the response, and the fits show no systematic
curvature; and the zero end of each fit lands on windows we measured directly
with no covert work at all (\cref{fig:marginal-dose}). We record the missing
clock control as a limitation of the platform (\cref{sec:discussion}).

\begin{figure}[t]
	\centering
	\includegraphics[width=\columnwidth]{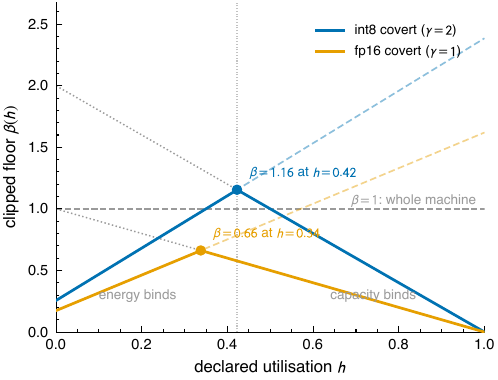}
	\caption{The relaxed bound $\beta(h)$
	(\cref{eq:betafloor}) from the jointly feasible fp16-declared bands,
	for covert int8 ($\gamma = 2$, blue) and covert fp16 ($\gamma = 1$,
	orange), both priced at Experiment~3's marginal lower limits. Faint
	dashed lines are the unclipped energy-only widths $\beta_{\mathrm{en}}(h)$
	($\beta_{\mathrm{en}}(1) = 2.39$ and $1.62$): energy is the binding constraint left of
	each crossing, capacity right of it, and both bind at $h^{*}$. Dotted
	lines are the capacity clips $(1-h)\,\gamma$; markers are the worst
	cases of \cref{eq:betaworst}.}
	\label{fig:betafloor}
\end{figure}

\paragraph{Assembling $\beta$.}
We now combine the three experiments to evaluate the relaxed bound $\beta$
(cf.\ \cref{tab:inputs}):
\begin{itemize}
	\item the declared band, from Experiment~1's fp16 cells:
	$[r_{\mathrm{lo}}^{\mathrm{dec}}, r_{\mathrm{hi}}^{\mathrm{dec}}]
	= [1.10,\,2.19]$\,pJ/op. This is narrower than the cross-format
	envelope of \cref{tab:inputs}, which pins no format: we assume
	the protocols of \cref{ass:declaration} let the verifier confirm the declared work
	ran in its declared format, so the declared edges are the fp16 cells'
	own;
	\item the overhead width, from Experiment~2:
	$\Delta P_0 = 41.2$\,W;
	\item the covert cost, from Experiment~3's marginal lower limits:
	$r_{\mathrm{lo}}^{\mathrm{cov}} = 0.51$\,pJ/op for covert int8
	($\gamma = 2$) or $0.76$\,pJ/op for covert fp16 ($\gamma = 1$);
	\item capacity $C_{\max}$ and $\gamma$ (\cref{eq:capratio}), from the
	two formats' published peaks rather than the rates our kernels
	achieve;
	\item the declared utilisation $h$: we take the worst case
	$\beta^{*} = \max_h \beta(h)$ (\cref{eq:betaworst}).
\end{itemize}
Together these give, with covert int8,
\begin{equation}
	\beta^{*} \;=\; 1.16
	\quad\text{at}\quad h^{*} = 0.42,
	\label{eq:betafloormeasured}
\end{equation}
and an energy-only width at full declared utilisation of
$\beta_{\mathrm{en}}(1) = 2.39$. With covert work held to fp16 the pair
is $\beta^{*} = 0.66$ at $h^{*} = 0.34$ and
$\beta_{\mathrm{en}}(1) = 1.62$. \Cref{fig:betafloor} traces $\beta(h)$
for both scenarios.

Three remarks. First, $\beta_{\mathrm{en}}(1)$ is an energy-only
diagnostic, not a physical floor: at full declared utilisation the
capacity clip $(1-h)\gamma$ is zero, so no covert work fits and
$\beta(1) = 0$. Second, the clip is evaluated at published peak
capacities rather than the rates our kernels achieve. Peaks give the covert
stream more room than it can really use, which is the adversary-favourable
direction: on our own measured peaks ($246.9$\,Top/s fp16 and $307.7$\,Top/s
int8, so $\gamma = 1.25$ rather than $2$) the headline would read
$\beta^{*} = 0.91$ rather than $1.16$.
Third, we can measure how much of this conceded width a real attack
occupies. On a second card of the same SKU we ran pairs of runs matched
in total energy: an honest run executing $N$ declared fp16 iterations on
expensive operands, and a cheating run executing the same iterations on
cheap operands plus a covert int8 fill sized to draw the same energy. Every
cheating run passes the no-alarm test of \cref{eq:noalarm} while
executing $0.19$--$0.41$ machines of covert work at $h = 0.10$--$0.216$.
The witness therefore establishes $\beta_{\mathrm{ID}}(0.216) \geq 0.41$,
against the relaxed model's $\beta(0.216) = 0.72$; the attack does
not exhaust the bound because no single schedule reaches every edge of
the band at once. Restricting the model to utilisations the attack can
occupy ($h \lesssim 0.24$) would lower the worst case to ${\approx}0.77$,
but that reflects our kernels rather than the adversary's best, so we do
not quote it as a bound. \Cref{app:witness} records the protocol.

\section{The capability ladder}
\label{sec:decreasing-beta}

The headline of \cref{sec:empirical-measurements}, $\beta^{*} = 1.16$
(\cref{eq:betafloormeasured}), is a relaxed worst-case bound under
\cref{ass:meter,ass:declaration} alone: the verifier reads the power and
checks the declared precision, and nothing else. In practice the verifier can
make a series of reasonable checks and assumptions to remove ambiguity and push
down $\beta^{*}$. Writing the
energy-only width of \cref{eq:betaen} with its inputs exposed,
\begin{equation}
	\beta_{\mathrm{en}}(h) \;=\;
	\beta_{\mathrm{en}}\bigl(h;\;
	r_{\mathrm{hi}}^{\mathrm{dec}},\,
	r_{\mathrm{lo}}^{\mathrm{dec}},\,
	r_{\mathrm{lo}}^{\mathrm{cov}},\,
	\Delta E_0\bigr),
	\label{eq:betadials}
\end{equation}
we see that the triple of \cref{eq:triple} together with the overhead width are the only
dials to turn to affect $\beta_{\mathrm{en}}$.

\emph{The declared edges $r_{\mathrm{hi}}^{\mathrm{dec}}$ and
$r_{\mathrm{lo}}^{\mathrm{dec}}$.} These are accessible to the verifier, at
least in principle, because they price work the prover has declared and the
verifier may re-run. The factorisation of \cref{eq:widthfactor} maps the available actions one can take to narrow that width: the declared precision is checked,
so $w_\kappa = 1$, while the remaining factors $w_\alpha w_C w_V$ provide open degrees of
freedom. Observations can remove the ($w_V$, $w_C$) freedom;
re-executing the declared work on the verifier's own hardware measures
$\bar r^{\mathrm{dec}}$ directly, bypassing $w_\alpha$ rather than pinning it.

\emph{The covert price $r_{\mathrm{lo}}^{\mathrm{cov}}$.} By definition, this is not accessible to the verifier; the prover still chooses the covert format, operands and locality freely,
and the meter never resolves them. However, a threat model or an assumption about what the
adversary is trying to do does introduce restrictions, for example
restricting covert work to a measured workload class, or assuming that we know
its numeric format.

\emph{The overhead width $\Delta E_0$.} Similarly, this dial is not
accessible to the verifier but can be restricted through reasonable
assumptions. For example, assuming that the card is executing throughout the
window being priced, so its overhead cannot be that of a cold card.

\begin{figure}[t]
	\centering
	\includegraphics[width=\linewidth]{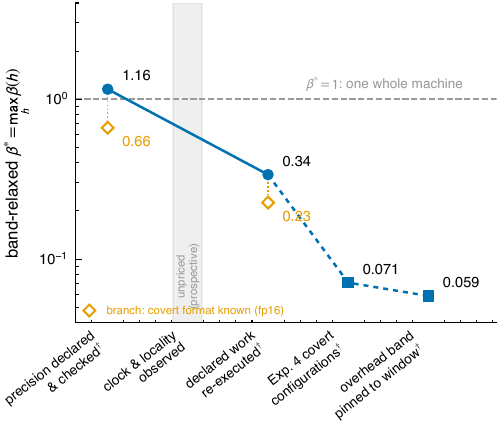}
	\caption{The capability ladder of \cref{tab:ladder} on a log scale;
	every plotted value is the relaxed worst case
	$\beta^{*} = \max_h \beta(h)$. Filled circles are capabilities the
	verifier gains; square rungs are forced by the threat model rather than
	granted by a channel; the shaded slot is the unpriced prospective
	clock/locality rung, and daggered labels carry the disclosed conditions of
	\cref{tab:ladder}. Hollow diamonds are the covert-format branch, dropped
	from the two rungs it re-prices: it is not a continuation of the chain, and
	does not compose with the rungs below. The $\beta^{*} = 1$ line marks one
	whole machine of conceded ambiguity.}
	\label{fig:ladder}
\end{figure}

The rest of this section takes these restrictions in turn. They compound. Each
one keeps everything granted above it and turns one more dial, so the bound can
only fall as we descend: from the headline $\beta^{*} = 1.16$, where the meter
and the declared format are all the verifier has, to a conditional
$\beta^{*} = 0.059$ at the bottom, with every restriction in place.
\cref{tab:ladder,fig:ladder} walk down that ladder.

\begin{table*}[t]
	\centering
	\small
	\begin{tabular}{@{}llcc@{}}
		\toprule
		verifier gains & factor acted on, and the edges it leaves & $\beta^{*}$ & $\beta_{\mathrm{en}}(1)$ \\
		\midrule
		precision declared \& checked$^{\dagger}$ & $w_\kappa = 1$; band $[1.10, 2.19]$ & $1.16$ & $2.39$ \\
		clock \& locality observed & $w_V, w_C \to 1$ \emph{(unpriced)} & --- & --- \\
		declared work re-executed$^{\ddagger}$ & \emph{bypasses} $w_\alpha$: band $\to$ gap & $0.34$ & $0.35$ \\
		measured Experiment~4 transformer blocks$^{\ddagger}$ & denominator: $0.51 \to 2.53$ & $0.071$ & $0.072$ \\
		overhead band pinned to the window$^{\ddagger}$ & $\Delta P_0$: $41.2 \to 31.3$\,W & $0.059$ & $0.059$ \\
		\midrule
		\multicolumn{4}{@{}l@{}}{\emph{branch --- covert format known to be fp16}$^{\S}$} \\
		\quad at the precision rung & $\gamma$: $2 \to 1$; denominator $0.51 \to 0.76$ & $0.66$ & $1.62$ \\
		\quad at the re-execution rung & \emph{as above}, band $\to$ gap & $0.23$ & $0.24$ \\
		\bottomrule
	\end{tabular}
	\caption{The capability ladder. Each row grants the verifier one more
	capability, cumulatively from the top, and moves one dial of
	\cref{eq:betadials}; the top row is the headline of
	\cref{eq:betafloormeasured}. Every row is the relaxed worst case
	$\beta^{*} = \max_h \beta(h)$ (\cref{eq:betaworst}) for the fp16-declared,
	int8-covert case study of \cref{sec:empirical-measurements}, computed
	directly from the edges beside it; rate edges are in pJ/op. We include
	$\beta_{\mathrm{en}}(1)$, the energy-only width of \cref{eq:betaen}, as an
	additional diagnostic. The clock and locality row is left
	unpriced, for the two reasons given in \cref{sec:decreasing-beta}.
	Disclosed conditions, inherited by every row below their first appearance:
	$\dagger$ --- the declared work is executed in its declared format;
	$\ddagger$ --- re-execution transfers at an observed operating point.
	$\S$ --- the branch restricts covert work to fp16 instead of int8, and is
	quoted only at the two rungs where its denominator is measured: it acts on
	the same dial as the measured-workload row, so the two do not compose
	(\cref{sec:decreasing-beta}).}
	\label{tab:ladder}
\end{table*}

\paragraph{Observe the clock and locality (unpriced; dial $r_{\mathrm{hi}}^{\mathrm{dec}}$).}
This rung would set $w_V \to 1$ and $w_C \to 1$, leaving the declared band at
its activity core $w_\alpha$ alone. Both are hardware-wide observables the power
meter does not measure. We deliberately quote no number for it, for two
reasons.

First, no verifier can do this today. Three routes might get there: the card
could attest its own clock telemetry, the facility could lock its clocks by
policy, or an external sensor could read the clock off the electromagnetic
signature (\cref{sec:discussion}). None of the three is currently deployable.

Second, and independently, we have not measured $w_V$. Neither
Experiment~1 nor the operand sweep of \cref{app:operands} sweeps the clock. In
both, the governor held it inside a $\pm3\%$ window. So the ${\approx}1.5$ we
carry for $w_V$ is \cref{sec:empirical-measurements}'s device-physics estimate,
not a measurement (\cref{tab:inputs}). 

\paragraph{Re-execute the declared work ($1.16 \to 0.34$, conditional; dial $r_{\mathrm{hi}}^{\mathrm{dec}}$).}
The verifier re-runs the declared work on its own hardware and measures what
it costs. It reads the operands that
actually ran, removing the prover's freedom to claim cheap data.

This collapses the declared band $r_{\mathrm{hi}}^{\mathrm{dec}} - r_{\mathrm{lo}}^{\mathrm{dec}} \to \epsilon$, where $\epsilon$ is a small but non-zero residual that quantifies the difference between the verifier's hardware and the prover's hardware. The verifier's trusted cluster is not the monitored facility, and energy per operation depends on the accelerator, its firmware, and its cooling. So the measured rate transfers only up to this hardware gap.

We measure the easiest version of the gap by re-measuring every band edge on three further A100s of the same SKU. The edges agree to $2$--$3\%$ across devices (\cref{app:bench}).
Same-SKU re-execution therefore transfers bands at the few-per-cent level.  However, across accelerator generations, firmware and cooling the gap is likely larger, and we have not measured it here. 

To move forwards we assume that $\epsilon$ is a few per cent of the
declared rate. We take $\epsilon \approx 0.05$\,pJ/op, informed by the same-SKU
spread but not measured across hardware, which gives $\beta^{*} = 0.34$ at
$h^{*} = 0.83$. We treat this as a sensitivity analysis in $\epsilon$ rather than
a measured bound: at the same-SKU spread itself ($0.02$--$0.04$\,pJ/op) it
reads $\beta^{*} = 0.29$--$0.32$; at our stated $0.05$\,pJ/op, $\beta^{*} = 0.34$; at a cross-vendor
$0.2$\,pJ/op, $\beta^{*} {\approx}0.54$. 

Moreover, even in the case where $\epsilon=0$, re-execution pays only at an
observed operating point, cf. the unpriced clock rung above. The verifier
measures the declared work at its own clock. If the prover's clock is unobserved
the prover may run the declared work anywhere on the DVFS curve. The transfer error is then not
the hardware gap but the whole DVFS factor, ${\sim}1.5$. So a verifier with only the meter, declarations and re-execution, but
no on-site sensing, stalls at $\beta^{*} \approx 1$. Re-executing the
declared work buys it almost nothing that checking the precision had not
already bought. Methods for pinning the clock are therefore a prerequisite for this rung, and no methods are currently deployable.

\paragraph{Restrict covert work to the measured transformer-block configurations
($0.34 \to 0.071$; dial $r_{\mathrm{lo}}^{\mathrm{cov}}$; Experiment~4).}
This is a workload-class restriction, granted by restrictions introduced by the threat model.
Covert work is unseeable by definition: no declaration reaches it, so the
unrestricted hardware edge remains the int8 marginal cost of
$0.51$\,pJ/op.  But the cheapest operations at that edge are operand zeros and in practice hidden zero-ops are no
threat as they compute nothing.  For this rung we instead restrict covert work to either of the
two int8-quantised transformer-block configurations measured in Experiment~4,
or to a class for which the verifier can independently justify the same
marginal-cost floor. We do not claim that every useful computation costs at
least this much, but we take it as a representative example of useful work beyond operand zeros.

We measured this by taking an int8 quantised transformer block built from one trained Pythia-6.9B
layer, run on real text activations rather than synthetic
patterns, with the quantisation scales applied. Every metered operation is
counted, both the int8 matmuls and the fp16 attention. We swept that block
as the covert dose inside Experiment~3's construction
(\cref{eq:doseslope}), varying only the number of covert blocks between
windows. We ran two shapes of the block, which differ in how much of the
operation count is fp16 attention rather than int8 matmul, and so bound the
per-operation cost from either side. Regressing window energy on covert nominal
operations gives a marginal cost of $2.54$ and $3.21$\,pJ/op for the two
shapes, both fits linear to $R^{2} > 0.999$. We carry the one-sided $95\%$ lower limits,
$r_{\mathrm{block}} = [2.53, 3.16]$\,pJ/op.

For these two configurations, using the lower edge of the measured band raises
the covert denominator from $0.51$ to $2.53$\,pJ/op: every slack joule buys
${\approx}5\times$ fewer operations than silicon-floor accounting would
concede. The relaxed bound drops to $\beta^{*} = 0.071$ under this restricted
scenario. Again we emphasise that this is a conditional case study, not a hardware-wide lower bound on
the cost of useful computation.

\paragraph{Pin the overhead band to the window being priced
($0.071 \to 0.059$; dial $\Delta E_0$; Experiment~2).}
Like the measured-workload rung, this rung is granted by restrictions introduced by the threat model. The band of \cref{eq:measuredoverhead} spans Experiment~2's
twenty-five idle windows.  Its low edge, $62.9$\,W, is the cold-card cell---the
device before any CUDA context exists. But in practice the card is not
cold. The declared burst runs for $h\,T$ and covert work fills the rest of the
window. The card is therefore executing from one end of the window to the other,
and its operands stay in memory throughout. A card in that state cannot draw a
cold card's overhead. Restricting
the band to the fifteen windows in which a context and operand pool were
resident lifts the low edge to $73.4$\,W. The high edge does
not move: the $96.2$\,W peak is the post-heavy-kernel transient, and the priced
window's declared burst \emph{is} a heavy kernel. With the same $\pm5\%$ meter
tolerance,
\begin{equation}
	P_0 \in [69.7,\, 101.0]\;\mathrm{W}, \qquad
	\Delta P_0 = 31.3\;\mathrm{W},
	\label{eq:pinnedoverhead}
\end{equation}
which is $0.76$ of the published width, so the rung falls from $0.071$ to
$0.059$. The restriction is not specific to the bottom of the ladder: at the
re-execution rung it reads $0.28$ in place of $0.34$, and at the headline
$1.12$ in place of $1.16$, where the declared band and not the overhead carries
the width.

Even after pinning, most of the bottom rung is overhead. The overhead term
$b = \Delta P_0/(r_{\mathrm{lo}}^{\mathrm{cov}}\,F_{\max}^{\mathrm{dec}})$ is
about $70\%$ of $\beta^{*}$ at the bottom rung ($0.040$ of $0.059$) and at the
re-execution rung ($0.195$ of $0.28$). Reporting $P_0$ directly, over a trusted
on-site channel rather than bracketing it, would leave only the meter tolerance,
$\Delta P_0 \approx 8.5$\,W, and a bottom rung of $\beta^{*}=0.030$.

\paragraph{Branch: know the covert format ($1.16 \to 0.66$; dials $r_{\mathrm{lo}}^{\mathrm{cov}}$ and $\gamma$).}

This restriction is enabled by the threat model. We assume that the covert numeric format
is known. This is a reasonable assumption:  if compute capacity is being stolen to train a frontier model, that training is highly likely to run in fp16 or bf16 with fp32 accumulation. int8 pre-training is not a viable route to the capability
an adversary wants.

Restricting covert work to fp16 moves two things. First, the covert
denominator rises from the int8 marginal edge to the fp16 one,
$0.51 \to 0.76$\,pJ/op, both measured in Experiment~3
(\cref{sec:empirical-measurements}). That is a factor ${\approx}1.5$ on the declared and
the overhead term alike, since both carry $r_{\mathrm{lo}}^{\mathrm{cov}}$ in
their denominator. Second, the capacity ratio falls, $\gamma = 2 \to 1$: covert
operations no longer retire at twice the declared rate, so the clip becomes
$1-h$. At the headline rung the two together give $\beta^{*} = 0.66$. Carried down to the re-execution rung the branch reads $0.23$ in
place of $0.34$. \Cref{fig:betafloor} traces both scenarios.

\section{Discussion}
\label{sec:discussion}
We have shown how a power meter constrains hidden compute only to an identified
set. For a meter that also checks the declared precision, matched schedules
establish $\beta_{\mathrm{ID}}^{*} \geq 0.41$, while the relaxed bound
gives $\beta_{\mathrm{ID}}^{*} \leq \beta^{*} = 1.16$. The capability ladder
can lower the relaxed bound to $0.059$, but only under all of its stated
conditions: the declared operating point and locality are observed,
re-execution transfers with the assumed $0.05$\,pJ/op gap, covert work is
restricted to Experiment~4's two measured transformer-block configurations,
and the overhead is that of a resident card.

\paragraph{Limitations.}
The above claim on the value of $\beta$ has a series of caveats as follows. (i) The empirical values of \cref{sec:howbig,sec:decreasing-beta} come from one vendor, one
GPU model, one driver, and one cluster. Consequently the quoted values of $\beta$ are
A100-specific, though the framework and closed form of \cref{sec:floor} are general.
(ii) The covert denominator behind the headline,
$r_{\mathrm{lo}}^{\mathrm{cov}} = 0.51$\,pJ/op, is measured, but on synthetic
operands: its cheapest cell is int8 on zeros, which compute nothing.
Experiment~4 measures only two Pythia transformer-block configurations on real
text activations. The bottom rungs rest on the lower edge of that measurement;
it is not a universal lower bound for useful computation. (iii) The hardware gap behind the
re-execution rung (${\approx}0.05$\,pJ/op) is an assumption. The same-SKU
piece is grounded: band edges agree to $2$--$3\%$ across three A100s
(\cref{app:bench}). However, the cross-hardware gap (different
accelerators, firmware, cooling) remains unmeasured. (iv) The platform denied clock locking and so DVFS was never
swept. Accordingly the voltage factor $w_V \approx 1.5$ of
\cref{eq:widthfactor} rides on literature numbers rather than our own. The
declared band was therefore sampled with the governor
holding the clock inside a $\pm 3\%$ window, whereas the threat model gives
the prover the clock: a prover free to slide DVFS reaches a wider declared
band and possibly a cheaper covert floor, raising both $s$ and $b$ in
\cref{eq:betaen}. (v) Energy was read from on-die telemetry, whereas in
practice the verifier reads a retrofitted meter on the power feed. The on-die
counter carries no published GA100 accuracy specification, so the $\pm 5\%$
tolerance we apply is adopted from NVML's documented Fermi/Kepler figure, and
the ${\sim}100$\,ms power-telemetry cadence is an empirical characterisation
\citep{yang2023parttime}, not a vendor specification. That retrofitted meter
carries its own calibration band, and it also sees conversion and cooling draw
the die counter never reports. Both widen $\Delta P_0$, and so widen $\beta$. (vi) We measured one device, whereas a facility is
many devices behind shared power delivery and cooling. We do not offer a
facility-level, multi-device version of \cref{eq:betaen}. Each
device carries its own capacity clip and its own utilisation. The
prover chooses how the declared work is spread across devices and formats, so
the width of the identified set depends on an allocation the prover is free to
pick. How to make the multi-device bound precise is an open question.
(vii) The relaxed bound is a worst case in a strong sense. It
prices a window in which the declared work is at its cheapest, the overhead at
its lowest, and the covert work at its cheapest---all three at the same time.
We do not show that any single schedule can do all three at once. If none can,
$\beta_{\mathrm{ID}}$ is narrower than the $\beta$ we quote
(\cref{eq:relaxation}), and we have conceded the adversary
more than the silicon allows.

\paragraph{Richer telemetry.}
Throughout this work we have dealt exclusively with time-integrated power, i.e. energy. Because $r(t)$ and $P_0(t)$ may vary arbitrarily within their
bands, the time-resolved power trace carries no information about total compute beyond the integral. However, real hardware is not perfectly free and does not vary arbitrarily. Instead power-ramp limits, DVFS
governor dynamics, the instantaneous clip $F(t) \le F_{\max}$, and any
declared schedule the prover commits to all constrain how the trace
can move. A verifier who models these constraints can likely extract more from the
resolved trace than from its integral. The time-resolved trace also carries different information such as workload cadence or 
training-step periodicity. This could be useful for attacking different questions about what kind of compute
ran (e.g. training vs. inference) rather than how much; that question needs different machinery and we
do not treat it here.

Moreover, we have focused exclusively on a single analogue measurement channel: power. A verifier with richer
instrumentation or multiple simultaneous channels
(e.g. per-rail power, electromagnetic, thermal), cryptographically
authenticated telemetry (the protocols of \cref{ass:declaration}), or independently
controlled measurements the prover cannot shape (locked clocks,
tamper-resistant attested counters) would likely be able to further constrain the total covert compute. For example, take the clock. The power meter cannot see it, which is why
\cref{sec:decreasing-beta} leaves the clock-and-locality rung unpriced. An
antenna beside the rack might see it instead. A circuit switching at frequency
$f$ radiates at $f$ and its harmonics, and magnetic emanations from a GPU's
power delivery are measurable in practice \citep{maia2022magnetic}. A verifier
who could read the clock this way would get the voltage too, since that follows
from the hardware's $V$--$f$ table. The DVFS factor $w_V$ would fall to $1$, and
the declared band would narrow to its activity core $w_\alpha$ alone. We defer the study of multiple measurement channels to future work.

\section{Conclusion}
\label{sec:conclusion}

So, can power draw constrain covert compute? Only weakly on its
own. Against a meter that also knows the declared numeric format, matched
schedules demonstrate at least $0.41$ machines of physical ambiguity, while
the relaxed bound caps the worst case at $1.16$:
$0.41 \leq \beta_{\mathrm{ID}}^{*} \leq 1.16$.

However, the meter is rarely all a verifier has. Additional information and threat-model restrictions can narrow the relaxed
bound. If the declared operating point and locality are observed, re-execution transfers
with the stated hardware gap, covert work belongs to one of the two measured
transformer-block configurations (or a separately justified class with the
same cost floor), and the card has resident-workload rather than cold-card
overhead, then these restrictions bring $\beta^{*}$ to $0.059$. This is a case study of what those restrictions buy, not a universal
bound on arbitrary useful computation.

The lowering of $\beta^{*}$ brought by additional restrictions strongly motivates the case for reading power alongside other channels rather than by itself. The clock is the clearest example; the meter cannot see it, but an
antenna beside the rack might. A verifier who can read the clock fixes the voltage
too and narrows the declared band to its activity core. Pricing that rung,
replacing the on-die counter with a retrofitted meter on the power feed, and
carrying the bound from one device to a whole facility are natural next steps.

\ifdefined\anonymous\else
\section*{Acknowledgements}

We thank Anjay Friedman, Maximus Rafla, and Coby Joseph for
discussions and feedback that shaped this work.

TK and EG thank the Machine Alignment, Transparency \& Security (MATS) programme---its
directors, organisers, funders, and staff---who made it possible for us to
work on this project and who provided invaluable resources and support
throughout.
\fi

\section*{Declaration on AI usage}
Following the disclosure principles of the Leiden Declaration on Artificial
Intelligence and Mathematics \citep{alper2026leiden}, we record the AI tools
this work relied on. Large language models (primarily Claude Opus~5, with additional inputs from Claude Fable~5 and
GPT-5.6-Sol) were used for writing,
refactoring and debugging the benchmark and analysis code. The mathematics was derived without LLM assistance; LLMs were used to check the consistency of the algebra. Every reported number is regenerated from the committed records by scripts in the repository, and the authors take full
responsibility for the content of this paper.

\bibliography{references}
\bibliographystyle{icml2026}

\newpage
\appendix
\crefalias{section}{appendix}
\crefname{appendix}{Appendix}{Appendices}
\Crefname{appendix}{Appendix}{Appendices}
\onecolumn

\section{From the CMOS power law to the exchange rate}
\label{app:cmos}

This appendix derives the two models the body text adopts: the affine
forward map \cref{eq:forward} sketched in \cref{subsec:forward}, and the
factored exchange rate \cref{eq:cmos} of \cref{subsec:factors}. It closes
by collecting the measurements that support the affine map, and by showing
that the security bound needs less than it---a weaker, one-sided
condition---and what that condition demands of how each band edge is
measured.

Start from the power law. Expanding the
static term of \cref{eq:cmoslaw}, we split total chip power into a dynamic
(switching) part and a static part,
\begin{equation}
	P = P_{\mathrm{dyn}} + P_{\mathrm{stat}}, \qquad
	P_{\mathrm{dyn}} = \alpha\, C\, V^2 f, \qquad
	P_{\mathrm{stat}} = V I_{\mathrm{leak}} + P_{\mathrm{fixed}},
	\label{eq:powertotal}
\end{equation}
with $\alpha$ the activity factor, $C$ the total capacitance of the logic
(so $\alpha C$ is the capacitance switched per cycle), $V$ the
supply voltage, $f$ the clock, $V I_{\mathrm{leak}}$ the leakage power, and
$P_{\mathrm{fixed}}$ the cooling, I/O, and idle draw. The forward model
\cref{eq:forward}, $P = rF + P_0$, is a straight line in the operation rate $F$,
so the exchange rate is its slope, i.e. the marginal energy per operation, and the overhead
is the intercept,
\begin{equation}
	r = \frac{\partial P}{\partial F}, \qquad P_0 = P\big|_{F=0}.
	\label{eq:marginal}
\end{equation}
Note that one cannot simply plot measured power
against measured rate and read $r$ off the slope. The card does not have one
$r$: the cost per operation depends on what is executing (e.g. the format, the
kernel, where the operands live), and the choices that move a workload's
achieved rate $F$ also move its cost per operation $r$. Points collected across
workloads therefore lie on different lines through a shared intercept, and a
slope fitted through them averages over workloads rather than pricing any one
of them (Experiment~2's rate ladder shows exactly this: when the fit is opened beyond
the compute-bound shapes, its scatter grows sharply while its slope barely
moves). At a fixed workload, power \emph{is} linear in the rate and its slope
is $r$ (cf. \cref{eq:rdyn} below), but each workload carries its own value of $r$.

\paragraph{The dynamic term gives $r$.}
The operation rate is $F = N_{\mathrm{op}}\,f$, where $N_{\mathrm{op}}$ is
the number of nominal operations the datapath retires per cycle at the operating format. At a fixed
format,
$\partial f/\partial F = 1/N_{\mathrm{op}}$, and the exchange rate of
\cref{eq:marginal} follows by the chain rule, holding $\alpha$, $C$, and $V$
fixed,
\begin{equation}
	r \;=\; \frac{\partial P_{\mathrm{dyn}}}{\partial F}
	\;=\; \frac{\partial P_{\mathrm{dyn}}}{\partial f}\,\frac{\partial f}{\partial F}
	\;=\; \frac{\alpha\, C\, V^2}{N_{\mathrm{op}}}
	\;=\; \alpha\,\Bigl(\tfrac{C}{N_{\mathrm{op}}}\Bigr)\,V^2 .
	\label{eq:rdyn}
\end{equation}
Note that the clock cancels: $r$ depends on what each switching event costs
and how many operations it delivers, not on how fast the operations run. The
slope $r$ is constant in $F$, so $P_{\mathrm{dyn}} = rF$ is linear through the
origin, exactly the affine form the forward model \cref{eq:forward} assumes
and Experiment~2's rate ladder checks (\cref{sec:empirical-measurements}). The capacitance switched
per operation, $C/N_{\mathrm{op}}$, has two structural sources: the arithmetic
datapath, whose width is set by the numeric format, and the data-movement
path, set by where the operands reside and which kernel runs. Factoring it
accordingly,
\begin{equation}
	\frac{C}{N_{\mathrm{op}}} \;=\; \kappa(\text{precision})\;C_{\mathrm{eff}}(\text{locality},\,\text{kernel}),
	\label{eq:capfactor}
\end{equation}
with $\kappa$ collecting the format scale (datapath width and the operation
packing $N_{\mathrm{op}}$) and $C_{\mathrm{eff}}$ the locality and kernel
structure. Substituting \cref{eq:capfactor} into \cref{eq:rdyn} gives the
model \cref{eq:cmos}. We emphasise this is not a unique factorisation.
Instead it is a regrouping of the switching contributions by physical origin;
$\kappa$, $\alpha$, and $C_{\mathrm{eff}}$ are all switching quantities,
grouped by cause so that each factor corresponds to one property of the
workload that a verifier could separately check or constrain: the numeric
format, the operand data, and the memory locality and kernel.

\paragraph{Short-circuit sits in $r$, inside the switched capacitance.}
\citet{chandrakasan1992lowpower} also carry a \emph{short-circuit} term for
the direct-path current that flows while an input transitions and the NMOS
and PMOS networks briefly conduct together. It is triggered by the same
switching events as $P_{\mathrm{dyn}}$, so it scales with the operation rate and
survives the marginal derivative of \cref{eq:marginal}. In a well-designed
circuit it is a small fraction of the switching power, so we fold it into
the effective switched capacitance $C_{\mathrm{eff}}$ of \cref{eq:capfactor}:
it lands in $r$, but adds nothing new for a verifier to check.

\paragraph{Leakage sits in $P_0$, not $r$.}
The static power $P_{\mathrm{stat}}$ of \cref{eq:powertotal} flows whether
or not the machine computes, so
$\partial P_{\mathrm{stat}}/\partial F = 0$: by \cref{eq:marginal} it
contributes nothing to the exchange rate $r$ and is entirely the intercept
$P_0$, where the derivation of \cref{eq:integrated} already places it and
carries it as a bounded nuisance band. A leakage-per-operation term
$r_{\mathrm{leak}} = P_{\mathrm{stat}}/F$ would appear only if $r$ were
defined as the average energy per operation, $P/F$, which mixes in the
intercept and is operating-point dependent; we keep the marginal definition
\cref{eq:marginal}, consistent with \cref{eq:forward}, and so carry no
separate leakage term in $r$.

\paragraph{Is the affine map true?}
The derivation above gives the model theoretically; the measurements also
support it directly, in three places. Experiment~2's rate ladder is the
across-workload check: restricted to the compute-bound shapes, where like is
compared with like, metered power against sustained operation rate is linear
at $R^2 = 0.99$, and the fitted slope lands inside the fp16 exchange band
$[1.10,\,2.19]$\,pJ/op that Experiment~1 measures independently by total
quotient (\cref{sec:empirical-measurements}). Experiments~3 and~4 test the
covert term at fixed declared work: window energy is linear in the covert
dose to $R^2 > 0.999$ (\cref{fig:marginal-dose})---affine in the one
argument whose marginal cost the security bound uses. And the idle cells of
Experiment~2 pin the intercept directly (\cref{eq:measuredoverhead}). What
no measurement certifies is one affine map across all workloads at once:
as noted above, each workload carries its own slope, and the pooled ladder
fit degrades accordingly. So \cref{eq:forward} holds workload by workload:
within one workload power is a straight line in the rate, and different
workloads have different slopes. As the next paragraph shows, that is all
the bound needs.

\paragraph{The bound needs less than the affine map.}
\label{app:estimands}
Why is the workload-by-workload form enough? Because the security bound does
not, in fact, require one affine map across all workloads; it needs one
weaker, one-sided condition. To state it, write the true draw as
$P = g_x(F_{\mathrm{dec}}, F_{\mathrm{cov}})$, some device-dependent function
of what the silicon executes, with the subscript $x$ standing for the
execution details the verifier never sees---no linearity assumed. The
condition is then
\begin{equation}
	g_x(F_{\mathrm{dec}}, F_{\mathrm{cov}}) - g_x(F_{\mathrm{dec}}, 0)
	\;\ge\; r_{\mathrm{lo}}^{\mathrm{cov}}\,F_{\mathrm{cov}},
	\label{eq:incremental}
\end{equation}
for every admissible execution $g_x$: adding covert work at rate
$F_{\mathrm{cov}}$ on top of the declared work costs at least
$r_{\mathrm{lo}}^{\mathrm{cov}}$ joules per covert operation, however the
silicon runs it. The derivation
of \cref{eq:covmax} converts the honest energy slack into covert operations by
dividing by $r_{\mathrm{lo}}^{\mathrm{cov}}$---so under \cref{eq:incremental}
alone the bound holds, with no global affine map anywhere. The affine model is
the special case
$g_x = P_0 + r^{\mathrm{dec}}F_{\mathrm{dec}} + r^{\mathrm{cov}}F_{\mathrm{cov}}$,
which additionally makes $\beta_{\mathrm{en}}(h)$ linear in $h$.

The condition \cref{eq:incremental} dictates how each band edge must be measured. The declared edges
enter \cref{eq:covmax} only through their difference
$r_{\mathrm{hi}}^{\mathrm{dec}} - r_{\mathrm{lo}}^{\mathrm{dec}}$ at a common
saturated rate, so any baseline amortised equally into both edges cancels and
a measured total quotient $E/C$ is the correct estimand for the numerator (the
overhead enters once, separately, as $\Delta E_0$). The covert edge is
different: it must be a genuine one-sided lower bound on the incremental cost
in \cref{eq:incremental}, and a total quotient cannot be one, because a
quotient amortises baseline power (e.g. leakage, cooling, idle draw) that an
added covert operation does not pay. Every covert edge the bound uses is
therefore identified by a dose--response intervention (Experiment~3 for the
generic int8 edge, Experiment~4 for the useful-operand edge), never by a
quotient. The same logic is why \cref{tab:inputs} keeps
$r_{\mathrm{lo}}^{\mathrm{dec}}$ and $r_{\mathrm{lo}}^{\mathrm{cov}}$ distinct
even though they describe the same physical case, int8 on zeros. The marginal figure is the
cheaper of the two, and a cheaper covert denominator concedes the adversary
more operations per slack joule, so carrying it widens the bound---the safe
direction.

\section{The matched-energy witness protocol}
\label{app:boundkind}
\label{app:witness}

This appendix establishes two results that bracket the physical ambiguity.
The model bound $\beta^{*} = 1.16$ is an \emph{upper} limit: it puts each
band at its worst edge independently, and no single schedule is guaranteed to
reach all of those edges at once. A constructed cheating schedule, accepted by the meter, puts a
measured \emph{lower} limit of $0.41$ machines on the same ambiguity. The
physical identified set against a format-checked meter therefore obeys
\begin{equation}
	0.41 \;\leq\; \beta_{\mathrm{ID}}^{*} \;\leq\; \beta^{*} = 1.16,
	\label{eq:ambiguity-bracket}
\end{equation}
and this appendix records both
halves of that bracket.

\paragraph{The upper end: $\beta$ treats the bands independently.}
The maximisation that produced \cref{eq:betafloor} lets each of the three
bands range freely, so the prover may sit at
$r_{\mathrm{lo}}^{\mathrm{dec}}$, $E_{0,\mathrm{lo}}$ and
$r_{\mathrm{lo}}^{\mathrm{cov}}$ all at once. Nothing in the derivation
guarantees a single schedule reaches all three edges at once. For equality we must
assume that there exist admissible workloads attaining
$(\bar r^{\mathrm{dec}}, E_0, \bar r^{\mathrm{cov}}) =
(r_{\mathrm{lo}}^{\mathrm{dec}}, E_{0,\mathrm{lo}}, r_{\mathrm{lo}}^{\mathrm{cov}})$
jointly and under shared constraints: the \emph{same} observation window $T$,
the \emph{same} declared work $C_{\mathrm{dec}}$, and one device in a
\emph{common thermal and clock state} whose time the two streams share
at their \emph{achieved} rates. Treating the bands
independently is a relaxation of what a device can schedule, and
\begin{equation}
	\beta_{\mathrm{ID}}(h) \;\le\; \beta(h),
	\label{eq:relaxation}
\end{equation}
with equality only if one schedule can reach all three edges at once. Conceding the adversary more
than the silicon may allow is the safe direction for a security claim,
but it leaves the gap in \cref{eq:relaxation} unquantified. (One
bookkeeping rule keeps the bands honest: every number entering the bound
comes from one scenario a device could actually run---one declared format
and one covert format sharing one window. We never mix formats, for
example by taking the declared cost from one format's most expensive
cells and the covert capacity from another.)

\paragraph{The lower end: a schedule the meter accepts.}
Closing the gap from below takes a constructed schedule. The matched-energy
witness of \cref{sec:empirical-measurements} holds the trusted
declaration and the observation window fixed: run~A executes $N$ declared
fp16 iterations on expensive operands (the $r_{\mathrm{hi}}^{\mathrm{dec}}$
edge) and idles to a fixed $T = 10$\,s window; run~B executes the
\emph{same} $N$ iterations on cheap operands plus a covert int8 fill
tuned to match A's energy. The acceptance rule is one-sided: run~B is
accepted if its window energy stays below run~A's honest mean plus three
standard deviations of A's repeat scatter. At $h = 0.10$, $0.15$ and
$0.216$ the accepted covert load reaches $0.19$, $0.29$ and $0.41$
machines of declared-format capacity: measured, capacity- and
time-feasible lower bounds on the ambiguity. The match is tight: at
$h = 0.10$ the arms agree to $|\Delta E|/E = 0.9\%$ over eight
interleaved repeats ($1291.3 \pm 4.4$ against $1279.3 \pm 5.6$\,J), while
B hides $3.5\times 10^{4}$ covert int8 iterations behind the no-alarm
rule. And the witness is conservative: run~A sits at the card's software
power cap, which truncates its honest energy and tightens the bar B must
clear. The topmost witness is thin: at $h = 0.216$ B's most expensive
repeat stays below the threshold by only $0.86\%$ of the honest mean, so
$0.41$ is the largest load we could get accepted.

\paragraph{Why the two ends do not meet.}
The witnesses meet the shared constraints above (i.e. one window, one
declaration, one device, etc.) but reach
none of the band edges: the achieved rates are $1.84$\,pJ/op for the
expensive declared arm (band ceiling $2.19$), $0.82$ for the cheap arm,
and $0.54$ for the covert fill (marginal edge $0.51$). That is why the
accepted $0.41$ at $h = 0.216$ sits below the model's
$\beta(0.216) = 0.72$: specifically,
$0.41 \leq \beta_{\mathrm{ID}}(0.216) \leq 0.72$. This is the relaxation
gap of \cref{eq:relaxation},
measured rather than assumed. Time is the other reason. The clip
\cref{eq:capclip} converts idle seconds into covert operations at
published peak rates, but at the rates these kernels achieve,
run~B's work no longer fits the window above $h_{\max} \approx 0.24$,
while the headline maximiser sits at $h^{*} = 0.42$; confined to
$h \le h_{\max}$, the model bound reads ${\approx}0.77$. We do not quote
$0.77$ as a bound, because the ceiling is set by our kernels' achieved
rates rather than the adversary's best. The general lesson survives the
caveat: a verifier reading time as well as energy constrains the
adversary more than the energy channel alone does, and joint energy--time
inversion is future work.

\section{Benchmark protocol, hardware, and provenance}
\label{app:bench}

This appendix is the measurement record: the hardware and provenance of
every experiment (including which card measured which band), the workload, how each swept driver is set, and
how the energy is metered. It is organised around Experiment~1
(\cref{sec:empirical-measurements}), which reads the total energy
quotient $E/C$ off a single A100 by holding the workload fixed and moving
one factor of \cref{eq:cmos} at a time. Experiments~2 and~5 reuse the same
protocol with a different swept axis, and Experiment~3 reuses its metering
inside fixed-length composite windows (declared burst + covert dose +
idle pad). The quotient is the band-edge estimand of
\cref{sec:empirical-measurements}---not the marginal rate
$r = \partial P/\partial F$ of \cref{eq:forward}, which Experiments~3
and~4 estimate as regression slopes---and we keep the two apart
throughout. Both declared band edges are read at the same saturated rate,
so their amortised baselines cancel in the band-edge difference the bound
uses; the ${\approx}5\%$ spread in achieved rate across cells leaves a
small residual, which widens the slack conservatively.

\paragraph{Hardware and provenance.}
Experiments~1--3 and~5 ran sequentially on one NVIDIA A100-SXM4-80GB (driver
595.84), in a single Slurm allocation ($340$, $85$, $70$, and $135$ recorded windows, respectively; these totals include the periodic reference-cell windows and the trimmed repeat of each cell, which the figure captions' window counts exclude).
Seven cards of
one SKU appear in total, by UUID prefix: \texttt{GPU-ac1cbae9}
(Experiments~1--3 and~5); \texttt{GPU-4bb93dfb}, \texttt{GPU-617f2717},
\texttt{GPU-aa412071} (cross-device sweep); \texttt{GPU-757ff02a} (the
trained-weight total quotient, reported only as a contrast to the marginal
edge); \texttt{GPU-747045fa} (Experiment~4, the incremental useful-work
edge that prices the bottom two rungs of \cref{sec:decreasing-beta}); and
\texttt{GPU-108c1549} (the matched-energy witnesses of
\cref{sec:empirical-measurements}). Experiment~4, the trained-weight
quotient, and the witnesses are therefore each read against declared edges
measured on a different card. The cross-device sweep bounds the error this
introduces: re-measured on three further cards of the same SKU, every band
edge agrees to $2$--$3\%$ (\cref{sec:decreasing-beta}). Full manifests accompany the released
records; the trained weight/activation tensors themselves are not
committed, but the manifests record their SHA-256 hashes and the GPU
pre-flight validation metrics, so the extraction is checkable against the
release but not replayable from the repository alone.

\paragraph{Operating point.}
The platform denied clock locking and power capping, so DVFS is a recorded
axis rather than a controlled one: each window logs the governor-selected
SM clock (achieved $1335$--$1410$\,MHz on the band sweeps), the bands are
stratified into $45$\,MHz clock strata, and no window was flagged with a
\emph{harmful} throttle (thermal or hardware power brake). The cells with
the most expensive operands do sit at the card's software power cap by design, which
truncates their measured power and is therefore conservative for a
most-expensive-rate edge.

\paragraph{Meter accuracy and cadence.}
\texttt{nvmlDeviceGetTotalEnergyConsumption} exposes a cumulative
millijoule register integrated in firmware. NVML publishes no accuracy
specification for this counter on GA100; the documented $\pm 5\%$ applies
to the power readings of the earlier Fermi and Kepler generations
\citep{nvidia_nvml}, and we
adopt that figure as a working meter tolerance wherever one is needed,
always in the direction that widens the bound. On the A100, NVML's power
readings update every ${\sim}100$\,ms and average only a $25$\,ms window
of each period \citep{yang2023parttime}. Whether the energy counter shares
this sampling is not documented. Each measured cell therefore
fills a ${\sim}1.5$\,s window (${\gg}100$\,ms), as described under
\emph{Metering the energy} below.

\paragraph{The workload and its exact operation count.}
Every cell runs a dense square matrix multiply $A\,B$ with
$A\in\mathbb{R}^{M\times K}$, $B\in\mathbb{R}^{K\times N}$ (we take
$M=N=K=n$). Each of the $MN$ output entries is a dot product of length
$K$---$K$ multiplies and $K{-}1$ adds---so by the standard
two-operations-per-multiply--add convention the matmul costs
\begin{equation}
	C \;=\; 2MNK \;=\; 2n^3 \quad \text{op},
	\label{eq:flopcount}
\end{equation}
counted as nominal operations of whatever format the cell runs (int8
multiply--adds included; \cref{subsec:forward}),
known analytically from the shapes alone, with no profiler or
hardware counter involved. A measured window runs \texttt{iters} such
matmuls back to back, so its nominal compute is
$C_{\mathrm{win}} = 2n^3\,\texttt{iters}$ and the energy per operation is
the quotient $E_{\mathrm{win}} / C_{\mathrm{win}}$. Because $C$ is exact under the
nominal convention (the literal scalar count is $MN(2K-1)$, within
$0.03\%$ of $2MNK$ at the band cells' $K = 2048$), the quotient
inherits the meter's accuracy directly; and because it is intensive
(energy \emph{per} operation), cells of different size and \texttt{iters}
compare directly.

\paragraph{Setting the swept drivers.}
The bound reads only the extremes of the quotient, so what
the sweep needs from each factor of \cref{eq:cmos} is its range. Every swept axis below is built to drive the factor it controls to both extremes.
Each must also move without disturbing the operation count
\cref{eq:flopcount}, or the quotient $E/C$ would change for two reasons at
once.
\begin{itemize}
	\item \textbf{Precision $\kappa$} is the tensor \texttt{dtype}:
	\texttt{fp32}, \texttt{tf32} (fp32 operands routed through the tf32
	tensor-core path), \texttt{fp16}, \texttt{bf16}, and \texttt{int8} (an
	integer-matmul kernel, int8$\times$int8$\to$int32). Lower precision
	charges less capacitance per multiply-accumulate, lowering $\kappa$.
	\item \textbf{Locality $C_{\mathrm{eff}}$} is operand residency
	at a byte-identical shape, not problem size: sweeping it by changing $n$
	would change the operation count and confound the comparison, so instead
	we fix the shape at $n = 2048$ and vary how many distinct operand sets
	the window rotates through iteration to iteration. The A100 has a
	40\,MB L2; at $n = 2048$ an fp16 $A$--$B$ operand set is ${\sim}17$\,MB,
	so:
	\begin{itemize}
		\item \emph{resident} (one operand set, reused every iteration): the
		working set fits the L2, so most traffic is served on-chip and HBM
		activation is avoided: low $C_{\mathrm{eff}}$;
		\item \emph{streamed} (eight operand sets, round-robined): a reuse is
		evicted before it recurs, so every iteration streams fresh tiles from
		HBM, engaging its arrays, peripheral logic, and I/O: high
		$C_{\mathrm{eff}}$.
	\end{itemize}
	Both arms run the identical shape, kernel, and operation count
	\cref{eq:flopcount}; only the memory-traffic path differs, so locality no
	longer confounds the count. The kernel stays compute-bound in both
	arms (a $2048^3$ matmul is high arithmetic intensity) so streaming
	raises $C_{\mathrm{eff}}$ without making the workload memory-bound, and
	\cref{fig:exchange-band} accordingly shows locality moves $r$ only
	slightly. The residency margin is comfortable only for the two-byte
	formats. An fp32 or tf32 operand set stores four bytes per element, so it
	is twice the size, ${\sim}34$\,MB against the $40$\,MB L2: its resident arm
	marginally spills, and the locality contrast we measure for those two
	formats is attenuated accordingly.
\end{itemize}
The third driver, supply voltage $V^2$ (DVFS), we could not sweep: the
platform denied user clock locking, so the clock was left to the governor
and merely recorded per run. The operand \emph{values} are a fourth axis:
Experiment~1 crosses all six operand distributions
with precision and locality in a full factorial, so the pooled envelope
already contains the operand spread; Experiment~5 re-runs the operand axis
alone, at fixed precision and locality, as the descriptive operand-only map
of \cref{app:operands}.

\paragraph{Metering the energy.}
Energy is read from the on-die NVML total-energy counter: the window energy
is one subtraction $E_{\mathrm{end}} - E_{\mathrm{start}}$, with a \texttt{cuda.synchronize()} fence
bracketing each latch so the delta spans exactly the window's kernels.
(The harness falls back to integrating the power query in a background
sampler where the counter is unavailable; no recorded window used the
fallback.) Each cell fills a ${\sim}1.5$\,s window with matmuls
run back to back on pre-allocated buffers, with no host traffic inside the
window: long enough that counter quantisation is negligible, and saturated
enough that the joules are the datapath doing arithmetic, not allocation or
transfer. A few warm-up windows absorb JIT and clock ramp, several measured
windows follow, and the per-cell median is kept; shuffling the cell order
and periodically re-measuring a fixed reference cell guards against slow
thermal drift. The largest per-cell median over the sweep is
$r_{\mathrm{hi}}^{\mathrm{dec}}$ with no format pinned, and the largest within
one format is that format's own ceiling;
\cref{fig:exchange-band} plots the full set.

\section{How Experiment~5 varies the operands}
\label{app:operands}

This appendix documents Experiment~5, the operand-only sweep: what it
measures and how each of its six operand distributions is constructed. The spread it measures,
$w_0 = 1.83$, independently supports Experiment~1's operand width
$w_\alpha = 1.88$, although no security bound rests on any number in this
appendix. The sweep also resolves the activity factor into named
distributions, which is what the threat-model rungs of
\cref{sec:decreasing-beta} act on when they exclude zero operands.

Experiment~5 holds the kernel, shape,
and precision (fp16) fixed and moves only the operand values. Every
configuration runs the identical matmul $A\,B$ with the identical operation
count \cref{eq:flopcount}, so the energy differences track the
switching-activity factor $\alpha$ of \cref{eq:cmos}---how many bits
physically toggle in the datapath as the multiply--accumulates stream
through. A datapath that sees the
same bits over and over charges little capacitance; one whose bits flip on
every operation charges the most. The six distributions are constructed to
walk $\alpha$ from its floor to its ceiling; they are the same six families
Experiment~1 crosses with precision and locality in its factorial
(\cref{app:bench}), so Experiment~1's fp16 cells contain this comparison as
one slice of a larger grid.

\begin{figure}[t]
	\centering
	\includegraphics[width=0.5\dimexpr\columnwidth-\columnsep\relax]{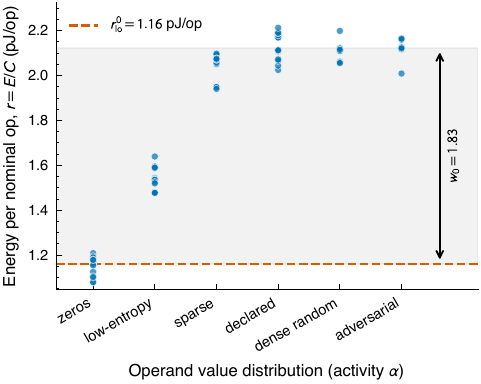}
	\caption{The shape of the activity factor $\alpha$ (Experiment~5, $n=96$
	windows across six operand distributions). Kernel, shape, and
	precision (fp16) are held fixed and only the operand values vary; the clock
	was governor-selected, not locked, and slides monotonically against operand
	cost, so this map is descriptive.
	The shaded span is the observed spread $w_0 = 1.83$, just below
	Experiment~1's matched-clock $w_\alpha = 1.88$, on the side the confound
	predicts. The
	minimum (zeros, dashed) is the descriptive fp16 operand floor
	$1.16$\,pJ/op as a total quotient---\emph{not}
	the covert denominator, which is Experiment~3's marginal edge.}
	\label{fig:operand-envelope}
\end{figure}

\paragraph{The measurement.}
Operand values alone spread $r$ by an observed factor
$w_0 = 1.83$ (\cref{fig:operand-envelope}), following the textbook ordering
zeros $<$ low-entropy $<$ structured-sparse $<$ dense-random $\approx$
declared-typical $\approx$ adversarial-toggle. The clock confounds the comparison.
Locking is privilege-denied (\cref{sec:empirical-measurements}), so the families
did not run at a common clock: the governor gave the cheapest (zeros, low-entropy)
the highest clocks and the most expensive the lowest, a monotone
$1410 \to 1335$\,MHz slide against operand cost. A faster clock carries a higher
supply voltage, and the voltage sets the energy per operation (\cref{eq:rdyn}),
so the slide squeezed the spread from both ends: the cheap families paid a
voltage premium that raised the floor, and the expensive families ran at a
voltage discount that lowered the ceiling. At one common clock both effects
disappear and the spread widens. The prediction is therefore directional:
$w_0 = 1.83$ should sit below a matched-clock measurement of the same factor.
Experiment~1 is such a measurement. Its operand families share a median
$1410$\,MHz and its width $w_\alpha = 1.88$ lies above $w_0$, on the predicted
side, agreeing to $3\%$. That $1.88$ is a median over ten format--locality
groups, only one of which runs what this sweep runs. Restricting to that
group, i.e.\ Experiment~1's fp16 streamed cells, gives $1.94$: again above $w_0$,
again on the side the confound predicts.

\paragraph{The distributions.}
Below we show each on a
$4\times4$ patch; the real operands are the same patterns at full size.

\paragraph{1. Zeros---the activity floor.}
Every entry is zero, so no bits ever change and $\alpha$ is at its minimum:
\begin{equation*}
	A_{\mathrm{zeros}} \;=\;
	\begin{pmatrix} 0 & 0 & 0 & 0 \\ 0 & 0 & 0 & 0 \\ 0 & 0 & 0 & 0 \\ 0 & 0 & 0 & 0 \end{pmatrix}.
\end{equation*}
This is the cheapest run at fixed fp16 and sets the descriptive fp16
operand floor, $1.16$\,pJ/op as a total quotient; combined with the int8
format in Experiment~1's factorial it reaches the cheapest measured quotient,
$0.786$\,pJ/op, which is $r_{\mathrm{lo}}^{\mathrm{dec}}$ with no format pinned.
(Both are declared entries. The covert entry
$r_{\mathrm{lo}}^{\mathrm{cov}}$ is Experiment~3's marginal cost, not either of
these quotients; \cref{sec:empirical-measurements,tab:inputs}.)

\paragraph{2. Low-entropy---a few distinct levels.}
Entries are drawn from only four values $\{0,1,2,3\}$. With so few distinct
bit patterns, few bits differ between neighbouring operands, so operand-bit
entropy, and hence switching, stays low:
\begin{equation*}
	A_{\mathrm{low\text{-}entropy}} \;=\;
	\begin{pmatrix} 1 & 0 & 3 & 1 \\ 2 & 2 & 0 & 3 \\ 0 & 1 & 1 & 2 \\ 3 & 0 & 2 & 1 \end{pmatrix}.
\end{equation*}

\paragraph{3. Structured-sparse---half the operands forced to zero.}
Dense random values, but a fixed half of the columns are deterministically
zeroed (every other column), giving exactly $50\%$ zeros in a regular
pattern: the structured sparsity real accelerators exploit, rather than
random dropout. The persistent zero columns roughly halve the switching
relative to a fully dense operand:
\begin{equation*}
	A_{\mathrm{sparse}} \;=\;
	\begin{pmatrix}
		0 & \phantom{-}0.4 & 0 & -0.9 \\
		0 & -0.7 & 0 & \phantom{-}0.2 \\
		0 & \phantom{-}0.8 & 0 & \phantom{-}0.5 \\
		0 & -0.3 & 0 & -0.6
	\end{pmatrix}.
\end{equation*}

\paragraph{4. Dense-random---generic busy data.}
Every entry is an independent uniform draw from $[-1,1]$: no zeros,
full-width mantissas, high but not maximal switching. This is the
realistic-workload reference:
\begin{equation*}
	A_{\mathrm{dense}} \;=\;
	\begin{pmatrix}
		\phantom{-}0.31 & -0.82 & \phantom{-}0.55 & -0.14 \\
		-0.67 & \phantom{-}0.09 & -0.48 & \phantom{-}0.73 \\
		\phantom{-}0.21 & \phantom{-}0.64 & -0.95 & \phantom{-}0.38 \\
		-0.52 & -0.27 & \phantom{-}0.86 & -0.41
	\end{pmatrix}.
\end{equation*}

\paragraph{5. Declared-typical---the realistic-activation stand-in.}
Every entry is an independent standard-normal draw, $\mathcal{N}(0,1)$: a
proxy for the activations and weights a genuine declared workload streams,
with full-width mantissas and no engineered structure. Its switching sits
alongside dense-random (i.e. high but not maximal) and it is the operand class
that characterises the honest-window energy scatter used by the declared-work
model:
\begin{equation*}
	A_{\mathrm{decl}} \;=\;
	\begin{pmatrix}
		\phantom{-}0.42 & -1.13 & \phantom{-}0.27 & -0.68 \\
		-0.35 & \phantom{-}0.91 & -1.42 & \phantom{-}0.11 \\
		\phantom{-}1.06 & -0.24 & \phantom{-}0.58 & -0.83 \\
		-0.72 & \phantom{-}0.39 & -0.16 & \phantom{-}1.24
	\end{pmatrix}.
\end{equation*}

\paragraph{6. Adversarial-toggle---the activity ceiling.}
Engineered for maximal switching: the matrix is a checkerboard of two
complementary bit patterns, $\texttt{0x5555}\ldots$ and its bitwise
inverse $\texttt{0xAAAA}\ldots$, so adjacent operands differ in every
bit (maximal Hamming distance) and the operand path sees maximal
input-level switching per operation. For floats the pattern is built in an
integer bit-view of the same width and reinterpreted as the float type, so
the raw bits are controlled exactly. Writing the
two patterns as $p$ and $\bar p$:
\begin{equation*}
	A_{\mathrm{adv}} \;=\;
	\begin{pmatrix}
		p & \bar p & p & \bar p \\
		\bar p & p & \bar p & p \\
		p & \bar p & p & \bar p \\
		\bar p & p & \bar p & p
	\end{pmatrix},
	\qquad
	p = \texttt{0x5555}\ldots,\;\; \bar p = \texttt{0xAAAA}\ldots
\end{equation*}
A matmul forms each output as a dot product
$\text{out}[i,j]=\sum_k A[i,k]\,B[k,j]$, streaming $k=0,1,2,\ldots$ and
loading one operand into the multiplier's input register each cycle.
Because every row and every column of the checkerboard alternates
$p,\bar p,p,\ldots$, that register walks
\begin{equation*}
	\underbrace{\mathtt{0101\,0101}}_{k=0}\;\to\;\underbrace{\mathtt{1010\,1010}}_{k=1}\;\to\;
	\underbrace{\mathtt{0101\,0101}}_{k=2}\;\to\;\cdots
\end{equation*}
flipping all its bits on every cycle. Dynamic CMOS energy is
paid per bit-transition, so this maximises the switching the operand
registers can drive: a $w$-bit operand toggles $w$ bits per cycle, against ${\approx}w/2$
for dense-random (two random words differ in about half their bits) and $0$
for zeros or any constant operand (the register never changes). The
checkerboard is simply the 2-D layout that makes the row sequence (feeding
$A$) and the column sequence (feeding $B$) both alternate this way at once,
for every output entry.

The energy-per-operation spread from the cheapest patch ($A_{\mathrm{zeros}}$)
to the densest operands (dense-random, declared-typical, and the adversarial
checkerboard), at fixed operation count, is the observed operand spread
$w_0$ of \cref{fig:operand-envelope}. Per
the clock confound above, the spread is
descriptive within-device evidence, not a certified population edge---and
maximal input-level toggling does not itself certify maximal switching
inside the tensor-core internals, whose operand encodings are opaque (nor
do zero operands mean no hardware bits switch); no security bound in the
paper rests on it.

\end{document}